\documentclass{article}

\usepackage{arxiv}

\usepackage[utf8]{inputenc} 
\usepackage[T1]{fontenc}    
\usepackage{hyperref}       
\usepackage{url}            
\usepackage{booktabs}       
\usepackage{amsfonts}       
\usepackage{nicefrac}       
\usepackage{microtype}      
\usepackage{lipsum}
\usepackage{float}
\usepackage{graphicx}
\graphicspath{ {./images/} }
\usepackage{subcaption}  
\usepackage{float}        
\usepackage{amssymb}
\usepackage[export]{adjustbox}
\usepackage{xcolor}
\usepackage{tcolorbox}
\usepackage{appendix}           
\usepackage{amsmath}
\usepackage{enumitem}
\usepackage{framed}
\usepackage{booktabs}  
\usepackage{tocloft}
\usepackage{multirow}
\usepackage{amsthm}
\usepackage{tikz}
\usepackage{caption} 
\usepackage{array}
\usepackage{tabularx}
\usepackage{booktabs}   
\usepackage{makecell}   

\renewcommand{\arraystretch}{1.4}
\usepackage[table]{xcolor} 
\definecolor{highlight}{RGB}{255,220,220} 

\usepackage{booktabs}
\usepackage{tabularx}
\usepackage{array}
\usepackage{makecell}
\usepackage[table]{xcolor}
\usepackage{threeparttable}
\usepackage{algorithm}
\usepackage{algpseudocode}
\usepackage{tcolorbox}
\tcbuselibrary{skins, breakable}

\newtcolorbox{promptbox}[2][]{
  enhanced,
  breakable,
  colback=#2!5!white,
  colframe=#2!75!black,
  fonttitle=\bfseries\sffamily,
  coltitle=white,
  attach boxed title to top left={xshift=4mm, yshift=-2.5mm},
  boxed title style={colback=#2!75!black, sharp corners},
  sharp corners=south,
  fontupper=\small\ttfamily,
  before upper={\setlength{\parskip}{0.5em}},
  #1
}

\usetikzlibrary{shapes.geometric, arrows.meta, positioning} 

\title{Exploring the Topology and Memory of Consensus: How LLM Agents Agree, Fragment, or Settle When Forming Conventions}

\author{
  Aliakbar Mehdizadeh \\
  Department of Communication \\
  University of California, Davis \\
  \texttt{amehdizadeh@ucdavis.edu} \\
   \And
  Martin Hilbert \\
  Department of Communication \\
  University of California, Davis\\
  \texttt{hilbert@ucdavis.edu}
}

\begin{document}
\maketitle

\begin{abstract}
How much should an LLM agent remember, and how should multi-agent systems be connected when trying to reach consensus? We show these two design choices interact in a way that flips the sign of memory's effect on coordination. Across 432 simulation runs of a networked Naming-Game on eight fixed 16-agent topologies, we vary memory depth and network structure. Longer memory slows the time to reach steady state in decentralized networks but accelerates it in centralized ones; the same parameter pushes the system in opposite directions depending on topology. Critically, "faster settling" in centralized networks means locking in to a fragmented plateau more quickly, not reaching system-wide consensus, which can be used to generate diverging opinions. We further document a memory-mediated speed–unity trade-off: centralized networks consistently preserve more competing conventions than decentralized networks, but their settling speed depends sharply on memory. At the agent level, within-network analyses show that high-betweenness bridges suffer a brokerage penalty while agents in locally clustered neighborhoods achieve higher coordination success. Finally, in search for analytically tractable generative mechanisms, we find that agents' choices are well captured by Fictitious Play, indicating belief-based rather than reward-based adaptation. The practical implication: memory depth and communication topology should be co-designed, not optimized in isolation.
\end{abstract}


\section{Introduction}

Driven by large language models (LLMs), generative AI agents~\cite{wang2024survey} function through iterative cycles of deliberation, coordination, and execution while interacting with their environment. Recent work shows that LLM-based agents can perform diverse real-world tasks beyond text generation. They can autonomously browse and interact with websites and computer interfaces~\cite{zhou2023webarena,deng2023mind2web}, execute multi-step planning through reasoning-and-acting frameworks such as ReAct~\cite{yao2022react}, perform long-horizon software-engineering tasks in agentic coding environments such as Claude Code, Codex, and Google Antigravity~\cite{anthropic2025claudecode,openai2025codex,google2025antigravity}, and control embodied or simulated environments such as SayCan~\cite{ahn2022can}. These systems demonstrate a transition from static language modeling to interactive, goal-directed, tool-using intelligence. As tasks become more complex, recent work has increasingly emphasized multi-agent architectures in which coordination, role specialization, and iterative communication can improve performance over single-agent systems~\cite{li2023camel,wu2023autogen,bo2024copper,tran2025multi}.

A critical class of solutions to recurrent coordination problems is the formation of social conventions~\cite{lewis1969convention,young1993evolution,bicchieri2005grammar,gelfand2024norm}. These arbitrary yet collectively adopted behaviors of agreement between agents, ranging from linguistic norms to moral frameworks~\cite{lewis1969convention,baronchelli2018emergence}, arise organically from local interactions without central authority~\cite{steels1995self,baronchelli2006sharp,centola2015spontaneous,hayek2020constitution,sugden1989spontaneous}. A pivotal question is whether LLM-based agent populations function similarly: Can they spontaneously develop universal conventions~\cite{werfel2014designing,brinkmann2023machine,wu2023autogen}? This question matters for both design and governance. If LLM agents can form conventions through local interaction, then memory and network structure become design variables that shape whether multi-agent systems coordinate reliably, fragment into local norms, or converge on system-wide agreement~\cite{dafoe2021cooperative}. 

In distributed multi-agent systems, network structure shapes how quickly information spreads and how robustly groups converge. Decentralized topologies can expose agents to diverse signals and support system-wide mixing~\cite{olfati2007consensus,granovetter1973strength}, whereas centralized structures can stabilize local agreement while insulating subgroups from one another~\cite{zhang2018fully,centola2007complex}. Memory adds a second design dimension: short histories support adaptation to recent interaction patterns, while longer histories can stabilize behavior but preserve obsolete information~\cite{xiong2025memory,salama2025meminsight,liang2026learning}. For LLM-based agents, coordination therefore depends on the joint design of communication topology and memory depth.

Prior work on cooperative AI emphasizes that artificial agents must be designed to find common ground under interaction constraints~\cite{dafoe2021cooperative}, while research on consensus in networked multi-agent systems shows that communication topology shapes convergence, robustness, and information flow~\cite{olfati2007consensus}. Recent LLM-agent frameworks such as CAMEL and AutoGen further demonstrate that multi-agent performance depends on how agents communicate, specialize, and coordinate~\cite{li2023camel, wu2023autogen}. From an engineering perspective, memory depth and network topology therefore function as control parameters: they can be tuned to favor rapid local coordination, broader system-wide consensus, or stable specialization depending on the application. The present study provides a controlled test of these parameters in a minimal coordination setting, offering design guidance for LLM-agent systems in which unmanaged memory or communication structure may lead to fragmentation rather than agreement.

Ashery et al.~\cite{ashery2025emergent} provide useful early evidence that populations of LLM agents, in well-mixed environments, can develop shared conventions through repeated coordination, highlighting the emergent nature of collective behavior in artificial agent systems. Their setting leaves two important extensions open. First, societies and multi-agent systems are rarely well mixed~\cite{newman2018networks}; interaction is usually constrained by social or communication networks. Second, agents vary in how much prior interaction history they retain and use~\cite{zhang2025survey,packer2023memgpt}. These two features, network structure and memory depth, are central to current LLM-agent design, but their joint effect on convention formation remains poorly understood.

In this study, we examine how network topology and memory depth jointly shape convention formation in LLM-agent populations. We place agents on the eight fixed 16-node network topologies introduced by Mason and Watts~\cite{mason2012collaborative}, varying their bounded memory length across $M \in \{2,5,10\}$. Agents play a repeated Naming-Game, a canonical model of convention formation in which local pairwise interactions can produce population-level agreement~\cite{baronchelli2018emergence,lewis1969convention,steels1995self,garrod1994conversation,wittgenstein2009philosophical}. The central challenge is equilibrium selection: how can a population converge on one convention among many equally viable alternatives without centralized coordination~\cite{lewis1969convention}?

We treat LLM agents as the system of interest in their own right. As autonomous AI agents become more pervasive~\cite{molinari2025towards,jiang2026humans,acharya2025agentic}, understanding how their populations coordinate is itself a substantive question, not a proxy for the analogous question about humans. Recent work has begun to document this directly: LLM-driven agent populations spontaneously generate biased network structures through preferential attachment and homophilic clustering on social attributes such as political orientation and religion~\cite{mehdizadeh2025homophily}, and individual LLM agents systematically update their opinions under peer pressure with model-specific conformity thresholds~\cite{mehdizadeh2025your}. The present study extends this line of work to the formation of arbitrary conventions on fixed topologies, asking how memory depth and network structure jointly determine whether populations of LLM agents converge or fragment.

\section{Methods}

To investigate our research questions, we developed an agent-based model simulating the well-known "Naming-Game" on a set of fixed social networks. Each agent in the model is an independent instance of an LLM, motivated to coordinate with its immediate neighbors.

\subsection{Agent-Based Model}

We adopt the canonical networked coordination paradigm: a pure-coordination Naming-Game~\cite{baronchelli2006sharp,baronchelli2018emergence,centola2015spontaneous} played on the eight 16-node, fixed-degree ($k=3$) topologies introduced by Mason and Watts~\cite{mason2012collaborative}. The benefit of choosing this setup is that its dynamics are well studied: it admits multiple absorbing consensus equilibria (one per available convention), exhibits symmetry-breaking convergence to a single global convention through purely local interactions, and shows convergence times that depend systematically on network structure (clustering and average path length)~\cite{dallasta2006nonequilibrium,baronchelli2018emergence}. These properties make it a well-characterized testbed for studying equilibrium selection, which is the phenomenon we wish to probe in LLM populations.

Each agent has a bounded memory of its last $M \in \{2,5,10\}$ interactions. These values represent short, intermediate, and relatively long memory windows. We deliberately exclude two limit cases: $M=0$, in which agents cannot condition on any prior interaction (which would isolate the prompt-only behavior of the LLM and is studied separately in our first-round bias check, Appendix ~\ref{sec:first_round_bias}); and unbounded memory, which would make early interactions permanently informative and confound memory depth with run length. Agents repeatedly choose from a shared set of ten arbitrary conventions: \texttt{C}, \texttt{D}, \texttt{F}, \texttt{J}, \texttt{M}, \texttt{P}, \texttt{Q}, \texttt{R}, \texttt{X}, and \texttt{Y}.

To check that these letters did not introduce strong intrinsic choice bias (for our purposes, these letters are arbitrary placeholders, while LLMs might be biased, given their training data), we analyzed first-round choices across $2{,}000$ independent no-memory prompts. Choices were close to uniform, and a chi-squared goodness-of-fit test did not reject uniformity ($\chi^2(9, N=2000)=4.35$, $p \approx 0.83$; Appendix~\ref{sec:first_round_bias}). This check helps ensure that subsequent convention formation reflects interaction dynamics rather than a strong initial symbol preference.

\paragraph{Interaction protocol.}
Edges in the 16-node networks are undirected. At each round, one edge is selected uniformly at random, and the two agents connected by that edge interact. Each agent simultaneously chooses one convention letter. If the two choices match, both agents receive $+100$ points; otherwise, both receive $-50$ points~\cite{schelling1960strategy,harsanyi1988equilibrium}. After the outcome is observed, each agent's score and memory are updated.

The simulation is an iterated game played over subsequent rounds. Agents use their memory of past interactions to adapt their strategy~\cite{fudenberg1998theory,macy1991learning}. The challenge thus is the problem of equilibrium selection~\cite{young1993evolution,kandori1993learning}. The central question becomes how a population of agents, through purely local interactions on a network, converges on one specific system-wide convention out of the many viable options~\cite{lewis1969convention,sugden1986economics,morris2000contagion,jackson2015games}.

Each simulation proceeded in discrete time steps, or rounds, with one round corresponding to a single pairwise agent interaction on a randomly selected network edge. A population cycle is defined as $16$ rounds, representing the expected number of interactions required for each agent to participate once on average. Simulations were run for a maximum of $1{,}600$ rounds, corresponding to $100$ population cycles; this horizon is consistent with prior Naming-Game studies on sparse networks of comparable size, which report convergence well within $\mathcal{O}(N^{1.4})$ rounds for $N=16$~\cite{dallasta2006nonequilibrium}, leaving enough margin to observe both transient and steady-state dynamics.

We initially scheduled $20$ independent runs for each of the $8 \times 3 = 24$ (network, memory) cells. A subset of runs did not complete due to transient API failures, which are unrelated to the experimental conditions (failures depend on API-side rate limiting, not on agent state). To preserve a fully balanced factorial design, we retained the largest common number of successful runs per cell (18), yielding 432 simulations. This sample size is sufficient to estimate per-cell mean convergence rates with standard errors below approximately $0.12$ on the $[0,1]$ success-rate scale (worst-case binomial bound), and the balanced design enables across-cell comparisons of network and memory effects.

\begin{algorithm}[H]
\caption{Networked Naming-Game with bounded-memory LLM agents}
\label{alg:simulation}
\begin{algorithmic}[1]
\State Initialize all agents with empty memory and zero score
\For{round $= 1$ to $1{,}600$}
    \State Pick a random edge of the network and identify its two agents
    \State Each agent independently queries the LLM with its current memory and chooses a convention
    \State Compare choices: if they match, both agents gain $+100$; otherwise both lose $-50$
    \State Each agent appends the interaction (own choice, partner's choice, success) to its memory; if memory exceeds $M$, the oldest entry is dropped
\EndFor
\end{algorithmic}
\end{algorithm}

The factorial design varies over the eight Mason--Watts topologies and $M \in \{2, 5, 10\}$, with 18 independent runs per cell, yielding 432 simulations and an interaction log of $432 \times 1{,}600 = 691{,}200$ dyadic decisions.

\paragraph{Agent Memory.}
Previous research has identified a non-monotonic relationship between memory length and system performance, suggesting a ``sweet spot'' where an \textit{intermediate memory length is optimal}~\cite{Horvath2012, Lu2018, Ma2021}. For instance, one experiment found the highest rate of human cooperation when subjects could remember the past two rounds~\cite{Ma2021}, while simulations have identified optimal ranges between 5--10 or 20--50 past interactions~\cite{Horvath2012, Lu2018}. This phenomenon can also be understood through an analogy to \textit{overfitting and underfitting} in statistical learning~\cite{hastie2009elements}.

Hence, memory can be expected to create a trade-off between adaptability and inertia. Short memory makes agents responsive to recent interaction patterns but vulnerable to noise, whereas longer memory stabilizes behavior but preserves obsolete information. Prior work on cooperation and repeated interaction similarly suggests that intermediate memory can be beneficial, with both very short and very long horizons producing distinct failure modes~\cite{Horvath2012, Lu2018, Ma2021}. We therefore use $M \in \{2,5,10\}$ to span short, intermediate, and longer bounded-memory regimes.

Following the bounded-memory tradition in repeated coordination games~\cite{juang2008learning,marsili2001learning} and the sliding-window short-term memory paradigm widely used in LLM agents~\cite{wang2024survey,park2023generative}, we implement memory as a fixed-size first-in, first-out deque. Each memory entry stores the agent's previous choice, its partner's choice, and whether the interaction succeeded. Before each new decision, this structured history is formatted as a natural-language prompt, allowing the LLM to condition its next choice on recent local interaction history.

\subsection{LLM Prompting}
\label{sec:llm_prompting}

The model is initialized with a detailed system prompt that is held constant across all API calls. This prompt specifies the rules, objective (score maximization), and required output format. We use the API's default decoding parameters at the time of the experiment (Gemini 2.0 Flash). Defaults may change with future model versions; the cross-model agreement check provides partial robustness against this. No parameters such as temperature, top-$p$, or \texttt{max\_tokens} were explicitly set, thereby relying on the default decoding behavior of the API for this model, which reflects typical deployment settings.

\begin{promptbox}[title=System Prompt]{blue}
You are a strategic AI agent participating in a multi-round coordination game. Your sole objective is to maximize your personal score.

\textbf{Game Rules:}

1. In each round, you will be paired with another agent.

2. You must choose a single convention (a letter) from a given list.

3. If your choice matches your partner's choice, you both succeed and earn 100 points.

4. If your choice does not match, you both fail and lose 50 points.

\textbf{Your Task:}

Analyze your recent interaction history provided in the prompt. Based on this history, decide which convention is most likely to result in a success in the next round.
\end{promptbox}

The user prompt for each decision consists of the agent's formatted memory and a randomly shuffled list of available conventions. Randomizing the list order reduces positional bias and ensures that prompts differ beyond the memory history itself. Below is an illustrative example of a user prompt an agent might receive.

\begin{promptbox}[title={User Prompt (illustrative example, $M=3$)}]{gray}
Here is the history of your last 3 interactions:

\hspace{1em}- Interaction 1: You chose: M, Partner chose: Y, Success: No

\hspace{1em}- Interaction 2: You chose: P, Partner chose: P, Success: Yes

\hspace{1em}- Interaction 3: You chose: P, Partner chose: J, Success: No

Your current score is: 500

Based on this history, which convention will you choose from the following list? [M, J, Q, Y, C, X, R, D, P, F]

Do not include any explanation or additional text.
\end{promptbox}

The prompt instructed the model to return only the convention letter and no explanation. Because memory length is the experimental manipulation, prompt length necessarily varies with $M$; this variation is part of the treatment rather than a nuisance parameter.

\subsection{Network Structures}
\label{sec:networks}

\begin{figure}[htbp]
\centering
\captionsetup{width=0.9\linewidth}
\includegraphics[width=0.9\textwidth]{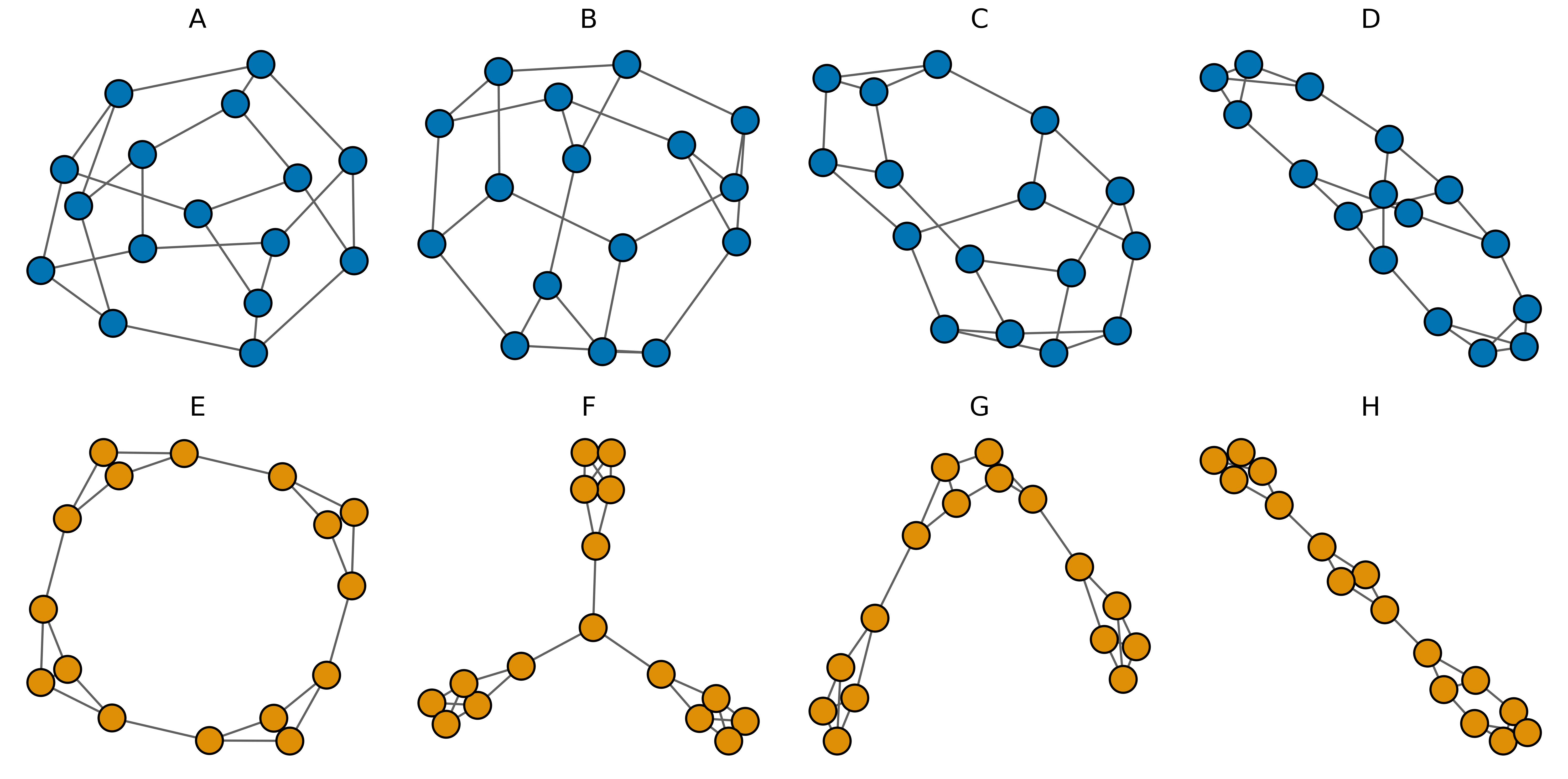}
\caption{The eight network structures used in the experiment, based on~\cite{mason2012collaborative}. Top row (A--D) represents decentralized networks, while the bottom row (E--H) shows centralized networks.}
\label{fig:networks}
\end{figure}

For this study, we adopt the eight network topologies first presented by Mason and Watts~\cite{mason2012collaborative}. Each network consists of 16 nodes where every node is connected to exactly three neighbors ($k=3$). These networks allow controlled manipulation of clustering and path length while maintaining constant degree, benefiting from prior known properties of these structures. Following the original study, we group the networks into two distinct categories based on their structural properties, as shown in Figure~\ref{fig:networks}. The specific structural properties for each network are detailed in Table~\ref{tab:networks_labeled}, which has been adapted from the original study.

\begin{itemize}
\item \textbf{Decentralized Networks (A, B, C, D):} These networks are characterized by a \textit{decentralized} structure, \textit{short average path lengths}, and \textit{low clustering coefficients}.
\item \textbf{Centralized Networks (E, F, G, H):} These networks feature more \textit{centralized} structures and/or significant \textit{local clustering}, which results in \textit{longer average path lengths}.
\end{itemize}

These topologies allow us to test whether communication structure affects the speed and completeness of convention formation.

\subsection{Behavioral Baselines}

To interpret the decision logic underlying the LLMs' observed choices, we compare their behavior to two adaptive baselines from the established behavioral game theory and learning literature, which act as external benchmarks: Fictitious Play (FP) and Reinforcement Learning (RL). These two rules are the canonical complementary baselines in models of learning in games and are widely used together to disentangle belief-based from reward-based adaptation~\cite{brown1951iterative,robinson1951iterative,erev1998predicting,camerer1999ewa,fudenberg1998theory}. Reinforcement-learning models have specifically been used to study the emergence of conventions in repeated coordination games on networks of human or simulated agents~\cite{nunner2022role}. The difference between them is that Fictitious Play, a belief-based rule~\cite{brown1951iterative,fudenberg1998theory}, picks the choice that best responds to what the partner is expected to do based on the partner's recent behavior, while Reinforcement Learning, a reward-based rule~\cite{erev1998predicting,sutton2018reinforcement}, picks the choice that has paid off best for the agent itself in recent memory. We also include a random-choice benchmark as a chance-level reference. 

\begin{table*}[htbp]
\centering
\caption{Behavioral baselines used to interpret LLM decision rules. The adaptive baselines differ in whether they use recent interaction history to form beliefs about partner behavior~\cite{brown1951iterative,fudenberg1998theory} or to reinforce previously rewarded actions~\cite{erev1998predicting,sutton2018reinforcement}. The random benchmark provides a chance-level reference.}
\label{tab:behavioral_baselines}
\small
\setlength{\tabcolsep}{6pt}
\renewcommand{\arraystretch}{1.25}
\begin{threeparttable}
\begin{tabularx}{0.95\textwidth}{p{2.2cm} p{2.4cm} p{3.0cm} X}
\toprule
\rowcolor{gray!12}
\textbf{Category} & \textbf{Baseline} & \textbf{Decision logic} & \textbf{Formalization} \\
\midrule
\multirow{2}{*}{\makecell[l]{History\\dependent\\adaptation}}
& \textbf{Fictitious Play (FP)}~\cite{brown1951iterative,robinson1951iterative,fudenberg1998theory}
& Forms beliefs about the partner's next move from empirical frequencies in memory and chooses accordingly.
& 
\[
U_t^{\mathrm{FP}}(i) = 100\, b_t(i) - 50\, [1 - b_t(i)]
\]
where $b_t(i)$ is the empirical probability, within the agent's memory window, that the partner selects convention $i$.
\\
& \textbf{Reinforcement Learning (RL)}~\cite{erev1998predicting,sutton2018reinforcement}
& Favors conventions that produced higher mean payoffs within the memory window.
&
\[
U_t^{\mathrm{RL}}(i) =
\begin{cases}
100\, \dfrac{S_t(i)}{n_t(i)} - 50\, \dfrac{F_t(i)}{n_t(i)}, & n_t(i) > 0 \\[6pt]
0, & n_t(i) = 0
\end{cases}
\]
where $S_t(i)$ and $F_t(i)$ are the numbers of successful and unsuccessful outcomes \emph{given} the agent chose convention $i$ within its memory window, and $n_t(i) = S_t(i) + F_t(i)$ is the number of times $i$ was chosen. Conventions never chosen receive the neutral utility $0$.
\\
\midrule
\textbf{Null benchmark}
& \textbf{Random}
& Selects uniformly across all available conventions, providing a chance-level reference.
&
\[
P(a_t = i) = \frac{1}{K}
\]
for $K$ possible conventions.
\\
\bottomrule
\end{tabularx}
\begin{tablenotes}[flushleft]
\footnotesize
\item FP and RL capture distinct forms of adaptive behavior: belief-based updating versus reward-based reinforcement. The two are widely used as complementary baselines in studies of learning in games and are special cases of the broader Experience-Weighted Attraction framework~\cite{camerer1999ewa}. Both utilities are expressed as expected per-action payoffs in the same units as the stage game and lie in $[-50, 100]$, ensuring that the FP and RL fits are directly comparable across memory sizes. Random defines the lower-bound reference.
\end{tablenotes}
\end{threeparttable}
\end{table*}

In both FP and RL, utilities are expressed in the same payoff units as the stage game. This makes the behavioral baselines directly interpretable: FP assigns higher utility to conventions that the partner is more likely to choose, while RL assigns higher utility to conventions that have produced higher realized payoffs for the agent. RL utility is the mean payoff \emph{per action}, normalizing by the number of times an action was chosen, $n_t(i)$, rather than by the memory length; this is the standard reinforcement-learning convention and makes utility reflect how well an action performed rather than how often it was tried. The distinction matters most at short memory, where most actions have been tried only once or twice. Following standard practice~\cite{camerer1999ewa,erev1998predicting}, the temperature parameter is estimated separately for each independent simulation run by maximum likelihood. We minimize the run-level negative log-likelihood with respect to $\tau$ over the bounded interval $[10^{-3}, 500]$ using Brent's method (\texttt{scipy.optimize.minimize\_scalar}, tolerance $10^{-6}$). The interval is wide enough that the optimum lies strictly in the interior for every run.

For FP and RL, multiple conventions can sometimes receive the same maximum utility. This is especially common early in the simulation, when agents have short or sparse histories. In these cases, requiring the model to select one arbitrary member of the tied set would understate its predictive performance. We therefore report tie-aware predictive accuracy: a prediction is counted as correct if the LLM's observed choice belongs to the set of maximum-utility conventions under the model. This rule affects only model evaluation; it does not affect the agents' choices during the simulation.

To evaluate probabilistic model fit, FP and RL utilities are converted into choice probabilities using a softmax rule, equivalent to the multinomial logit / quantal response specification widely used in behavioral game theory~\cite{mckelvey1995quantal,camerer1999ewa}:
\begin{equation}
P(a_t = i \mid h_t) =
\frac{\exp\left(U_t(i)/\tau\right)}
{\sum_{j \in \mathcal{A}} \exp\left(U_t(j)/\tau\right)},
\end{equation}
where $\tau > 0$ is the temperature parameter. Lower values of $\tau$ correspond to more deterministic best-response behavior, while higher values correspond to more diffuse choice probabilities. Following standard practice~\cite{camerer1999ewa,erev1998predicting}, the temperature parameter is estimated separately for each independent simulation run by maximum likelihood.

Model predictions are evaluated at the agent-decision level. Each dyadic interaction therefore contributes two prediction cases: one for Agent~1's choice based on Agent~1's memory prompt, and one for Agent~2's choice based on Agent~2's memory prompt. This avoids treating the dyad as a single decision-maker and allows each model to predict the actual information available to each agent at the moment of choice.

Note that utilities are computed as unweighted aggregates over the finite memory window. Hence, the models capture frequency-based (FP) and reward-based (RL) adaptation but abstract away from recency weighting, higher-order temporal dependencies, or strategic pattern recognition that richer learning frameworks such as Experience-Weighted Attraction~\cite{camerer1999ewa} would accommodate.

\subsection{Analytical Strategy}

The results are pursued systematically as a cumulative sequence. Section ~\ref{sec:EmpiricalConvergence} first establishes that repeated local interaction among LLM agents produces convergence toward shared conventions, identifying the basic phenomenon and revealing an early tension between local coordination and global agreement. Section~\ref{sec:dynamicfit} then grounds these macro-level dynamics in a micro-level account of choice by fitting our two adaptive baselines, Fictitious Play and Reinforcement Learning, to the agents' decisions, showing that individual choices follow systematic belief-based and reward-based regularities rather than arbitrary patterns. Section~\ref{sec:MemoryandNetwork} turns to the two manipulated design dimensions, memory depth and network topology, and analyzes their joint effect on local dyadic coordination. Section~\ref{sec:LocalCoordinationGlobalConsensus} builds on this by formally separating local coordination from global consensus, using steady-state fragmentation and settling time to surface a memory-mediated speed–unity trade-off between centralized and decentralized networks. Section~\ref{sec:structural_resilience} then shifts from the population level to the agent level, asking how structural position within a network shapes individual coordination success and thereby explaining where inside the topology the population-level patterns are produced. Finally, Section~\ref{sec:crossmodel} stress-tests the resulting account through a cross-model agreement check on sampled decision points, assessing whether the observed choice patterns generalize beyond the primary model. Each step earns the right to ask the next: convergence motivates the search for a decision rule, the decision rule supports the analysis of design dimensions, those dimensions sharpen the local-global distinction first noted in Section~\ref{sec:EmpiricalConvergence}, and that distinction is in turn explained by structural position within the network.

\section{Results}

\subsection{Empirical Convergence of Conventions}
\label{sec:EmpiricalConvergence}

\begin{figure}[htbp]
    \centering

    \begin{subfigure}[t]{0.48\textwidth}
        \centering
        \includegraphics[width=\textwidth]{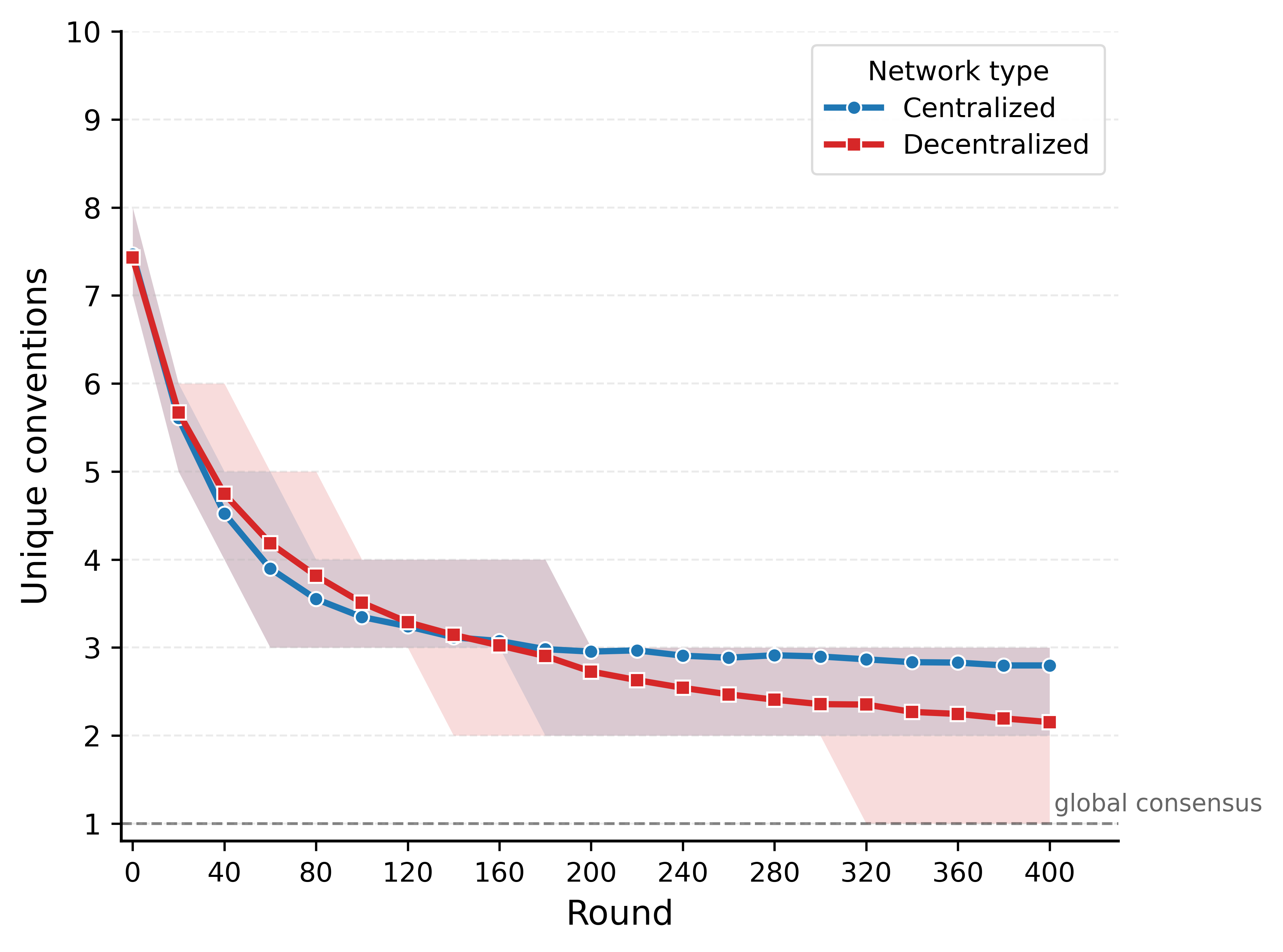}
        \caption{Mean convergence by network type.}
        \label{fig:empirical_convergence_network_type}
    \end{subfigure}
    \hfill
    \begin{subfigure}[t]{0.48\textwidth}
        \centering
        \includegraphics[width=\textwidth]{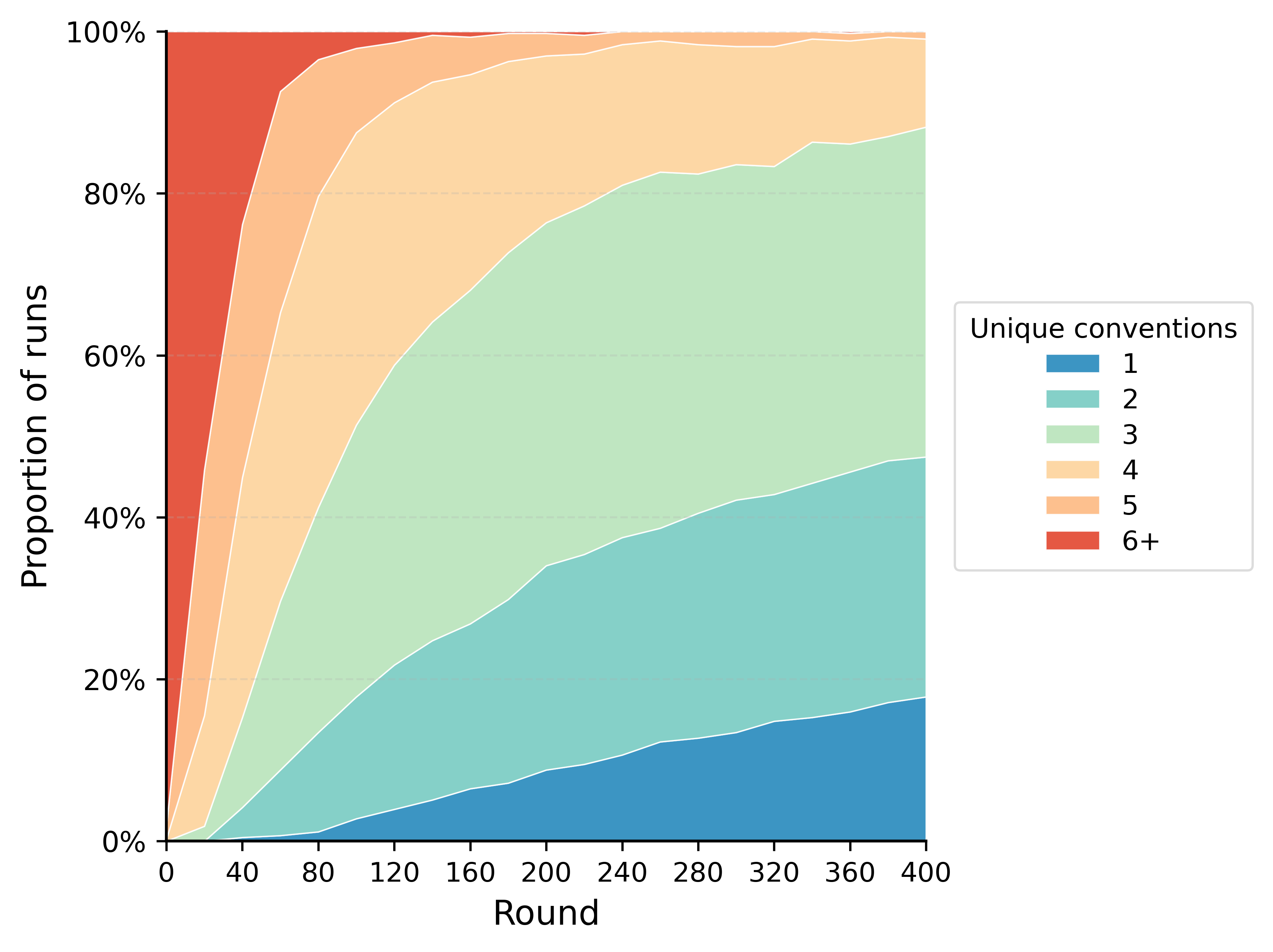}
        \caption{Distribution of convention diversity across runs.}
        \label{fig:empirical_convention_distribution}
    \end{subfigure}

    \caption{Early empirical convergence of conventions. 
    Panel (a) shows the mean number of unique conventions observed from round $0$ to round $400$, separated by network type and averaged across all memory sizes ($M \in \{2,5,10\}$). Shaded bands indicate the interquartile range across independent simulation runs. 
    Panel (b) shows the distribution of convention diversity across all independent simulation runs over the same time window. The stacked areas report the proportion of runs with one, two, three, four, five, or six or more unique conventions.}
    \label{fig:empirical_convergence_early}
\end{figure}

We first examine the raw coordination dynamics before introducing behavioral baseline models. The goal is to establish whether repeated local interaction among LLM agents produces population-level convergence toward shared conventions. Figure~\ref{fig:empirical_convergence_early} summarizes the early phase of this process from round $0$ to round $400$ using successive $20$-round windows.

Convention diversity declines rapidly across both network types, with the median run reaching three or fewer surviving conventions within roughly 200 rounds (Figure~\ref{fig:empirical_convention_distribution}). However, only a minority of runs reach a single global convention within this window.
Having established that conventions reduce over time, we next ask what decision rule produces this reduction at the agent level.

\subsection{Dynamic Model Fit}
\label{sec:dynamicfit}

Having established the empirical convergence pattern, we next examine whether agents' choices can be explained by simple adaptive models of repeated coordination. We compare two utility-based behavioral baselines: Fictitious Play (FP), which predicts choices from beliefs about partners' recent behavior, and Reinforcement Learning (RL), which predicts choices from the agent's own recent payoff history. A random-choice baseline provides a chance-level reference.

\begin{figure}[htbp]
    \centering

    \begin{subfigure}[t]{0.49\textwidth}
        \centering
        \includegraphics[width=\textwidth]{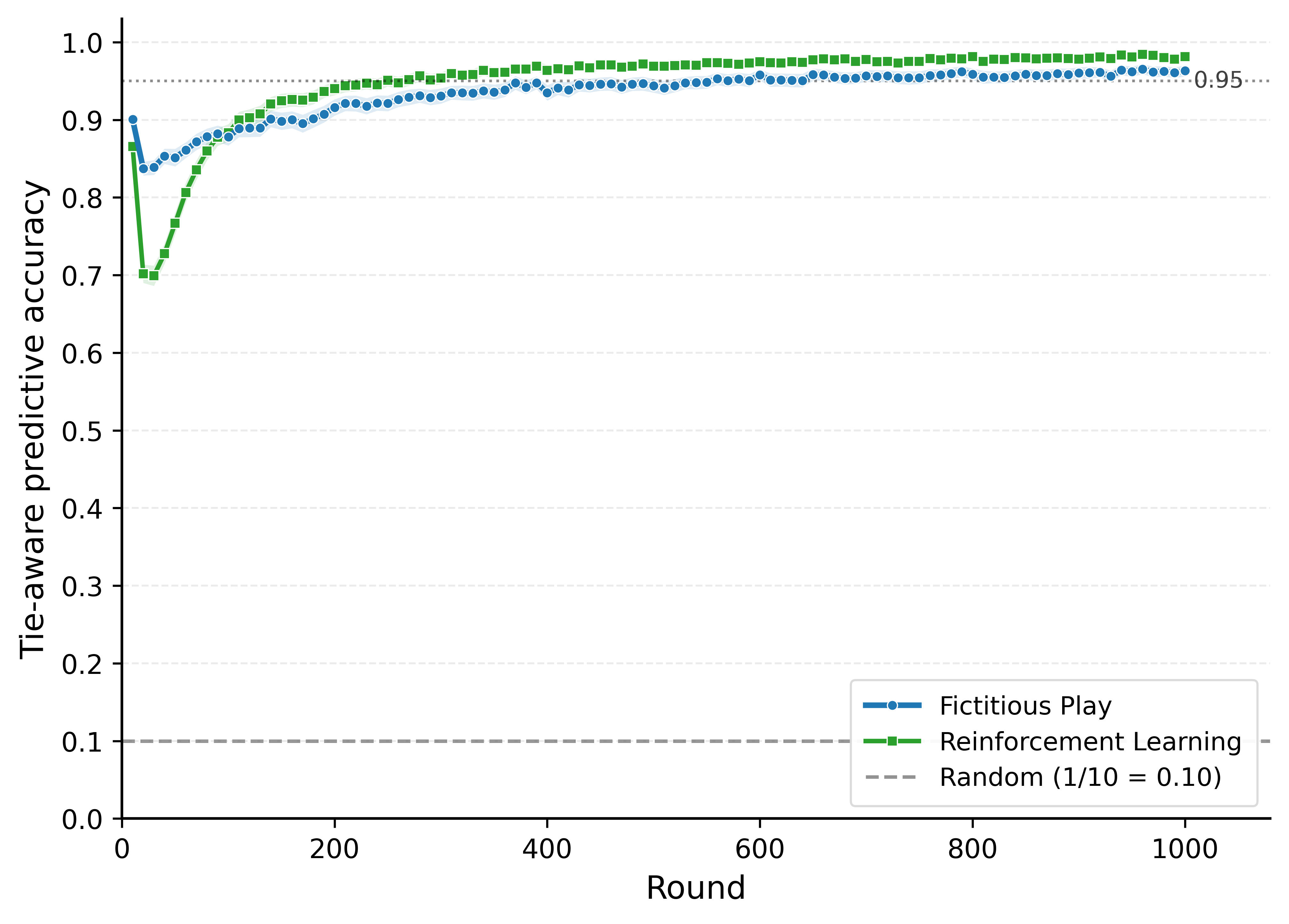}
        \caption{Predictive accuracy over time.}
        \label{fig:model_fit_over_time}
    \end{subfigure}
    \hfill
    \begin{subfigure}[t]{0.49\textwidth}
        \centering
        \includegraphics[width=\textwidth]{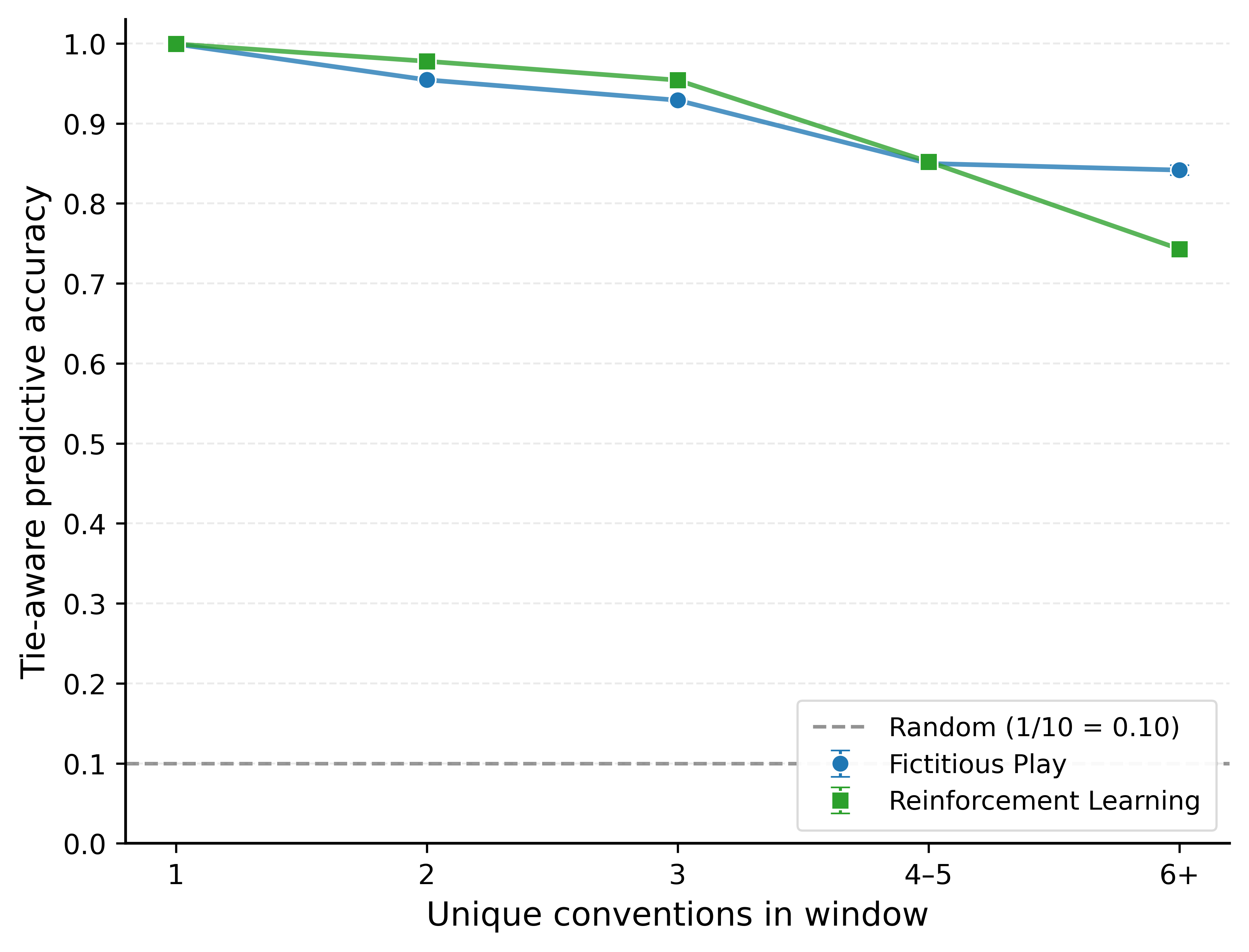}
        \caption{Predictive accuracy by convention diversity.}
        \label{fig:model_fit_by_fragmentation}
    \end{subfigure}

    \caption{Behavioral-model fit over time and across levels of convention diversity. 
    Panel (a) reports mean tie-aware predictive accuracy for Fictitious Play (FP), Reinforcement Learning (RL), and a random baseline over successive 10-round windows from round $0$ to round $1{,}000$. 
    Panel (b) reports tie-aware predictive accuracy conditional on the number of unique conventions observed in the same window. 
    FP and RL remain well above the random baseline even when several conventions remain active, indicating that their fit is not merely a consequence of late-stage consensus.}
    \label{fig:aggregated_analysis}
\end{figure}

The utility-based baselines provide a strong descriptive account of agents' choices. As shown in Figure~\ref{fig:model_fit_over_time}, both FP and RL rapidly outperform the random baseline. FP remains the strongest behavioral description over most of the observed trajectory, suggesting that agents' choices are especially consistent with belief-based adaptation to partners' recent behavior.

The high predictive accuracy is not simply a byproduct of late-stage convergence. Figure~\ref{fig:model_fit_by_fragmentation} shows that FP and RL remain substantially above the random baseline even in windows with high convention diversity. The models therefore capture regularities in the agents' local decision rules before the population has collapsed into a small number of conventions. More granular behavioral-model trajectories by network type and memory size are reported in Appendix~\ref{appendix:detailed_analysis} and Figure~\ref{fig:faceted_convergence}.

\subsubsection{Model Comparison Across Memory Sizes}

We next compare FP and RL across memory sizes using two run-level fit measures: negative log-likelihood (NLL) and the estimated softmax temperature $\tau$. NLL measures how much probability each model assigns to the agents' realized choices; lower values indicate better fit. Because predictions are evaluated at the agent-decision level, we report NLL per decision. The temperature parameter $\tau$ captures the stochasticity of the fitted softmax choice rule, with lower values indicating more deterministic predictions.

\begin{figure}[htbp]
     \centering
     \begin{subfigure}[b]{0.45\textwidth}
         \centering
         \includegraphics[width=\textwidth]{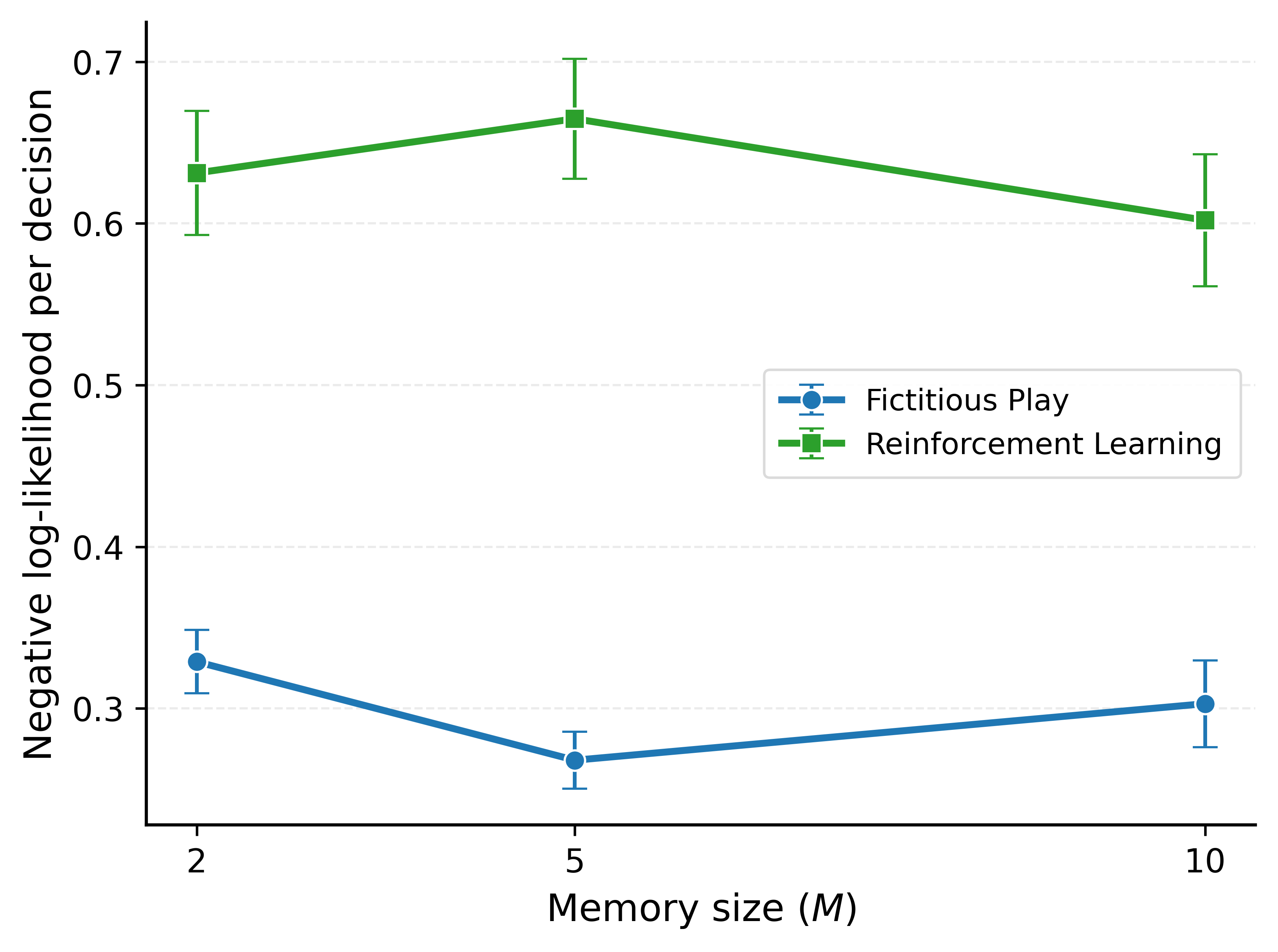}
         \caption{Model fit}
         \label{fig:nll_memory}
     \end{subfigure}
     \hfill
     \begin{subfigure}[b]{0.45\textwidth}
         \centering
         \includegraphics[width=\textwidth]{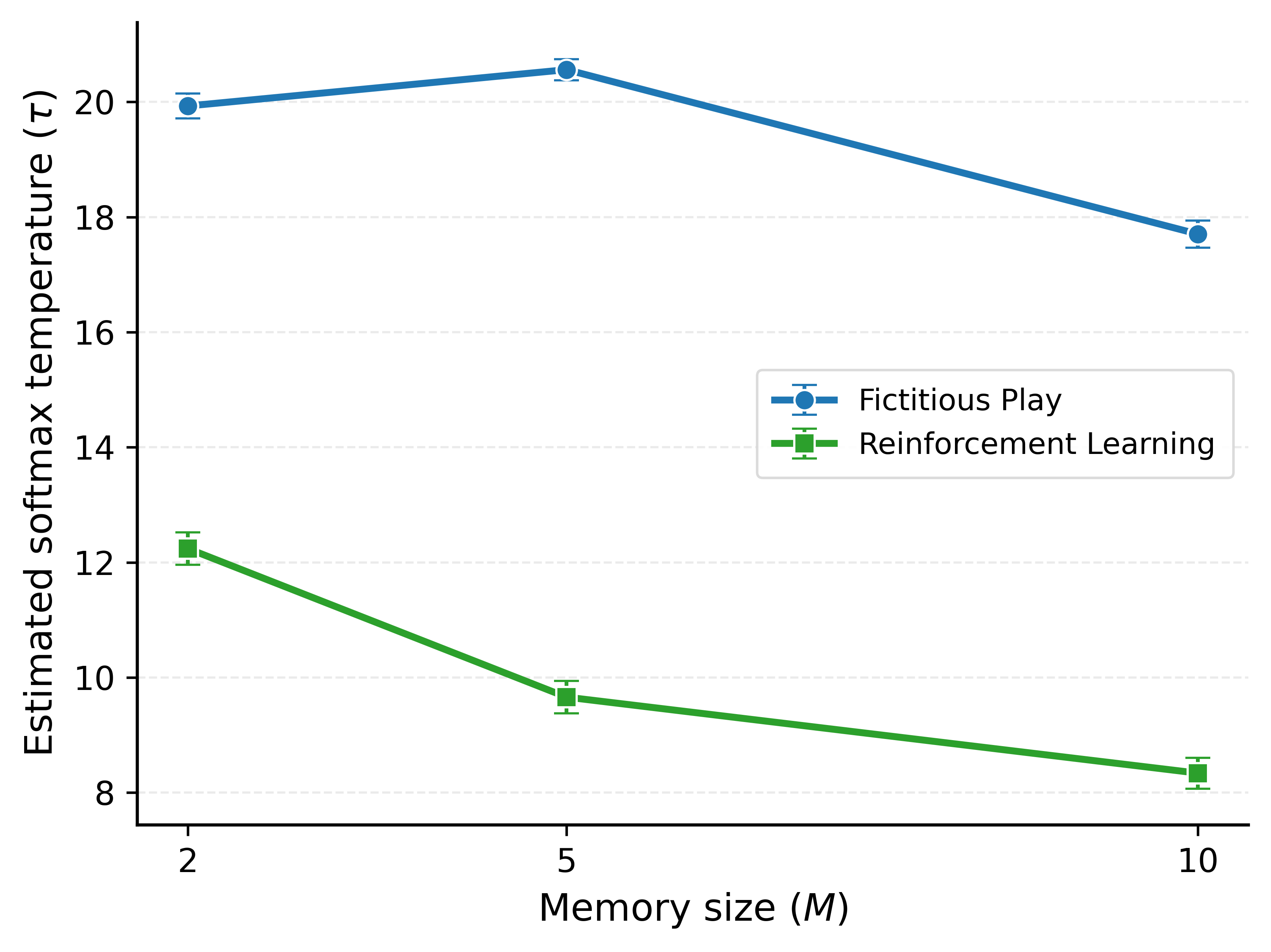}
         \caption{Decision temperature}
         \label{fig:tau_memory}
     \end{subfigure}
     
     \caption{Behavioral-model fit across memory sizes. 
     (a) Mean negative log-likelihood (NLL) per agent decision for Fictitious Play (FP) and Reinforcement Learning (RL); lower values indicate better fit. 
     (b) Estimated softmax temperature $\tau$; lower values indicate more deterministic choice behavior, while higher values indicate more diffuse choice probabilities. 
     Error bars indicate 95\% confidence intervals across independent simulation runs.}
     \label{fig:model_parameters_memory}
\end{figure}

Figure~\ref{fig:nll_memory} shows that FP achieves lower NLL than RL across all memory sizes. This indicates that agents' choices are better captured by belief-based expectations about partners' recent behavior than by reinforcement from the agents' own realized payoffs. FP's advantage over RL is largest at $M=2$, where the per-decision NLL gap is widest, and is smaller at longer memory, consistent with the per-action RL estimate being most volatile when few interactions per convention are remembered.

Although FP and RL achieve similar tie-aware accuracy after the early transient, their probabilistic fit differs. Tie-aware accuracy only asks whether the observed LLM choice belongs to the model's maximum-utility action set. NLL is stricter: it also measures how much probability the model assigns to the action the agent actually selected. FP therefore provides a better probabilistic fit across memory sizes, suggesting that the two models often identify similar high-utility action sets, but FP is better calibrated to the agents' realized choices, especially when convention diversity remains high. This belief-based pattern is consistent with concurrent evidence from a different LLM-agent setting: in opinion-change experiments, LLM agents update beliefs as a function of observed peer behavior with a response curve and model-specific conformity thresholds~\cite{mehdizadeh2025your}.

Figure~\ref{fig:tau_memory} reveals an informative tension: RL is fit with lower temperatures but achieves higher NLL than FP, meaning RL is confidently wrong more often than FP is cautiously right. FP's higher fitted temperature reflects appropriate uncertainty over a structured choice set, whereas RL commits to one option and pays a probabilistic penalty when the agent disagrees.

\subsection{Memory and Network Effects on Coordination}
\label{sec:MemoryandNetwork}

\begin{figure}[htbp]
    \centering

    \begin{subfigure}[t]{0.49\textwidth}
        \centering
        \includegraphics[width=\textwidth]{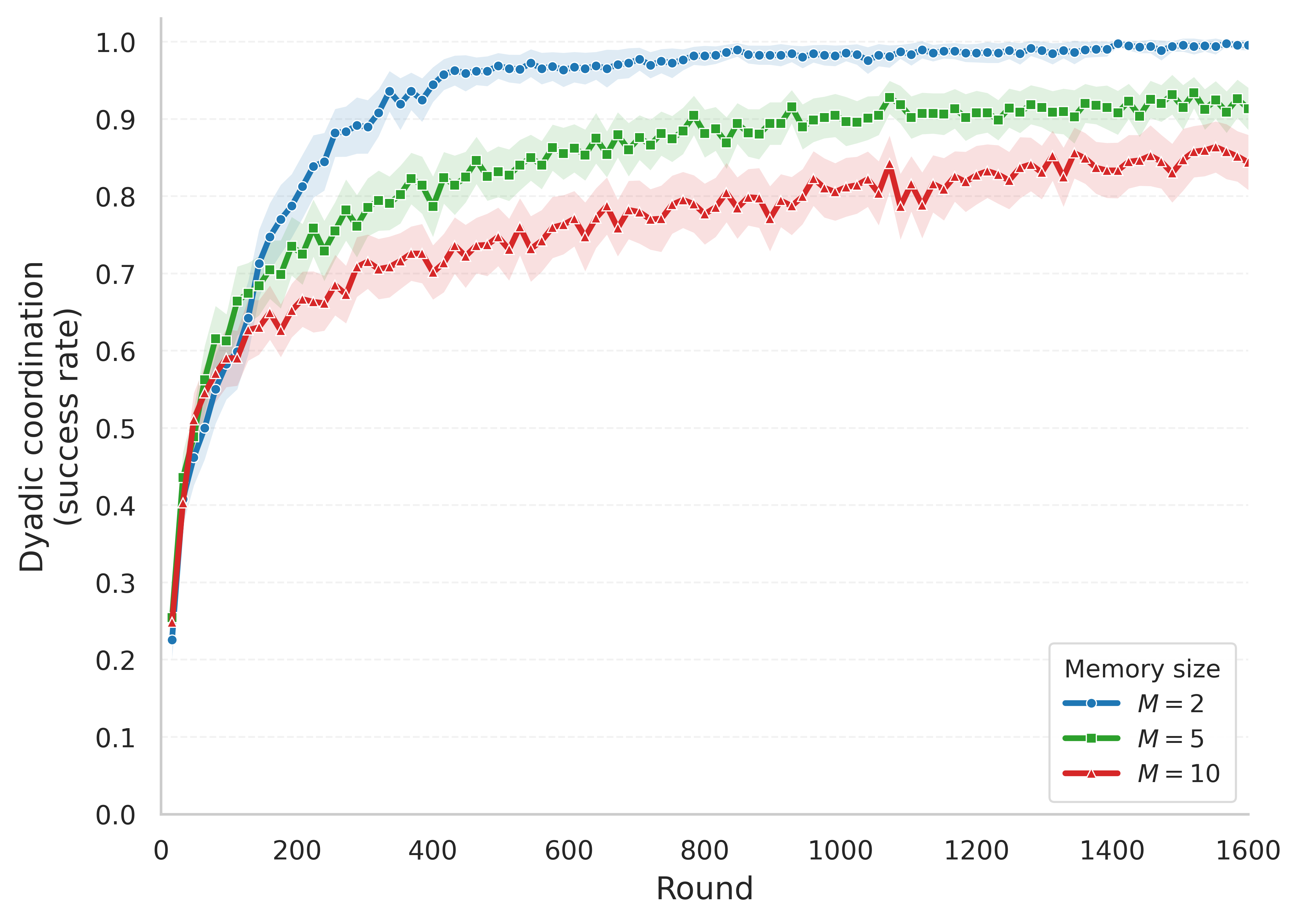}
        \caption{Decentralized networks}
        \label{fig:coordination_decentralized}
    \end{subfigure}
    \hfill
    \begin{subfigure}[t]{0.49\textwidth}
        \centering
        \includegraphics[width=\textwidth]{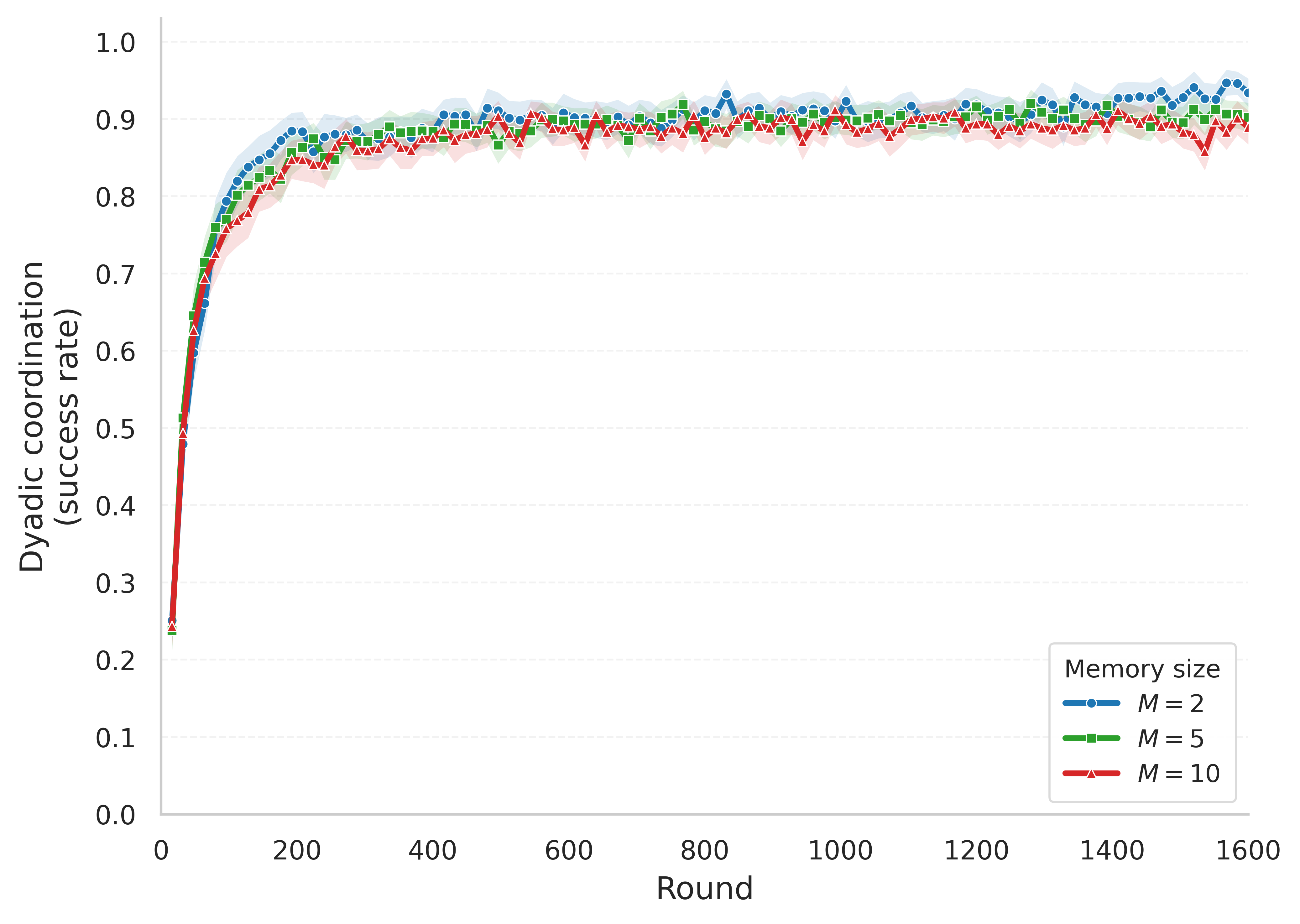}
        \caption{Centralized networks}
        \label{fig:coordination_centralized}
    \end{subfigure}

    \caption{Dyadic coordination over time by network type and memory size. Lines show mean pairwise success rates over simulation rounds for memory sizes $M \in \{2,5,10\}$. Shaded bands indicate 95\% confidence intervals across independent simulation runs.}
    \label{fig:faceted_convergence}
\end{figure}

We next examine how memory depth and network structure shape local coordination, measured as the dyadic success rate. Figure~\ref{fig:faceted_convergence} separates the trajectories for decentralized and centralized networks.

The effect of memory depends on network structure. In decentralized networks, shorter memory produces faster and higher coordination: agents with $M=2$ approach near-complete dyadic coordination, while agents with $M=10$ converge more slowly and remain at a lower success rate. This pattern is consistent with memory-induced inertia, in which longer histories preserve obsolete or locally conflicting information; we present direct evidence for this mechanism at the agent level in section ~\ref{sec:structural_resilience}.

In centralized networks, the memory effect on dyadic coordination is much weaker. All memory conditions converge to a high coordination plateau (0.89--0.94), and the trajectories remain close throughout the simulation. Centralized structures appear to provide strong local reinforcement that decouples \emph{pairwise} success from individual memory depth. As Section~\ref{sec:LocalCoordinationGlobalConsensus} shows, however, this apparent insensitivity does not extend to convention dynamics: memory has a large interactive effect on settling time and steady-state fragmentation in centralized topologies. Local pairwise success and population-level convention dynamics respond differently to the same parameter.

A two-way ANOVA was conducted to evaluate the effects of network type and memory depth on final dyadic coordination. The sampling unit for all inferential tests was the independent simulation run. The dependent variable was the per-run mean success rate over the final $50$ rounds of each simulation. Using run-level aggregates avoids treating dependent agent-level observations within the same simulation as independent. The analysis revealed a significant main effect of memory, $F(2,426)=53.76$, $p<0.001$, $\eta_p^2=0.202$, indicating that memory depth strongly affects final coordination success. The main effect of network type was not significant, $F(1,426)=1.17$, $p=0.281$, $\eta_p^2=0.003$. However, the interaction between memory and network type was significant, $F(2,426)=12.12$, $p<0.001$, $\eta_p^2=0.054$, showing that the effect of memory depends on network structure.

In decentralized networks, final coordination declined sharply as memory increased, from $0.996$ at $M=2$ to $0.915$ at $M=5$ and $0.852$ at $M=10$. In centralized networks, the decline was weaker, from $0.942$ at $M=2$ to $0.906$ at $M=5$ and $0.891$ at $M=10$. Thus, longer memory reduces coordination most strongly in decentralized networks, whereas centralized structures partially buffer the effect of memory depth. These results show how memory and topology shape local pairwise success. However, high dyadic coordination does not necessarily imply that the population has converged on a single shared convention.

\subsection{Local Coordination versus Global Consensus}
\label{sec:LocalCoordinationGlobalConsensus}

We next distinguish local coordination from global consensus. Local coordination is measured by the dyadic success rate, whereas global consensus is measured by the number of unique conventions remaining in the population. These two outcomes need not move together: agents can coordinate successfully within local neighborhoods while the population as a whole remains fragmented across multiple conventions.

\begin{figure}[htbp]
    \centering
    \includegraphics[width=0.68\textwidth]{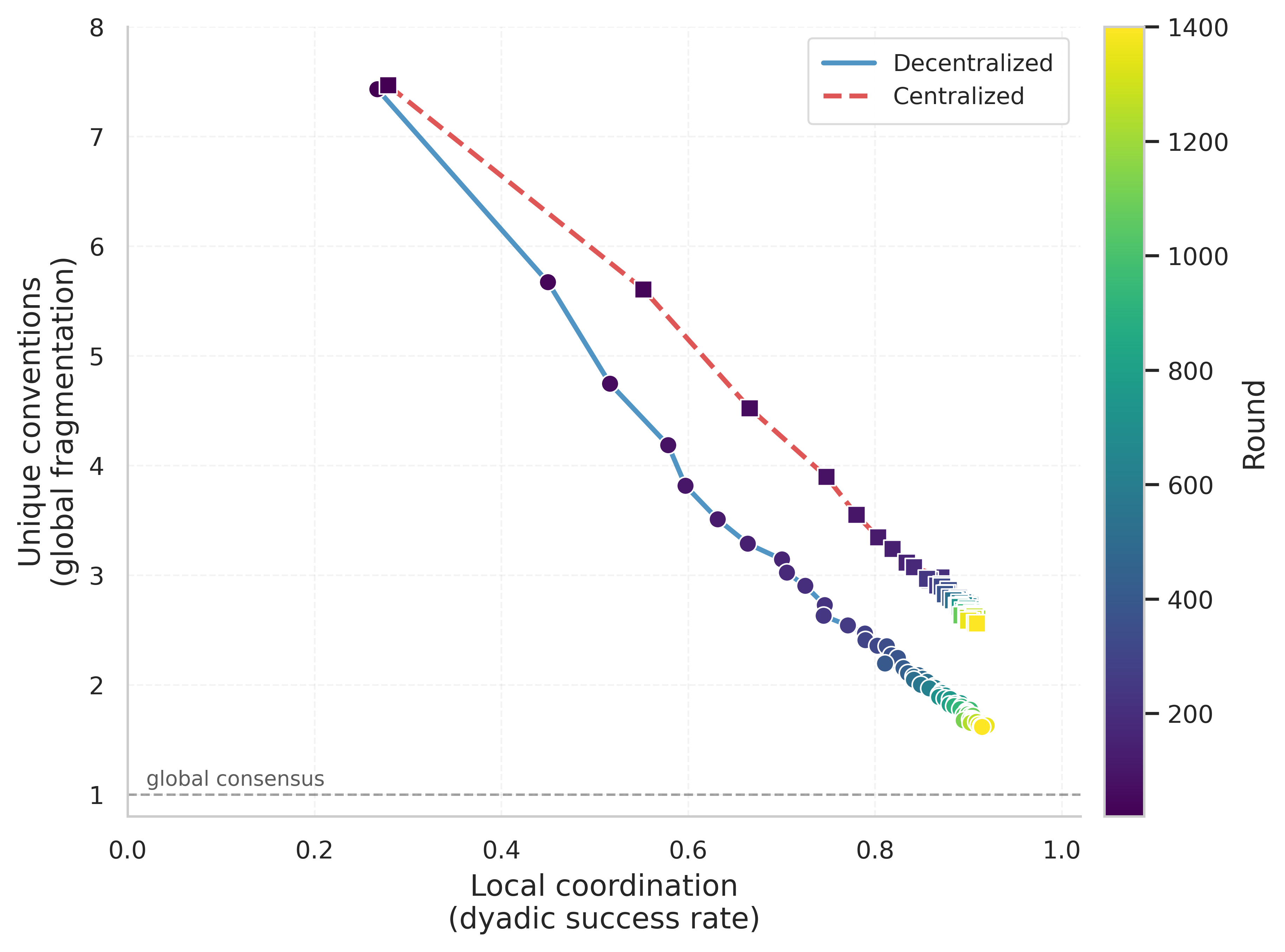}
    \caption{Local coordination versus global fragmentation. Each point represents a successive 20-round window, averaged across independent simulation runs and all memory sizes. The x-axis reports local dyadic coordination, measured as the success rate of pairwise interactions, while the y-axis reports global fragmentation, measured as the number of unique conventions observed in the same window. Color indicates simulation time, and line style separates centralized from decentralized network groups. The dashed horizontal line marks the single-convention global-consensus benchmark.}
    \label{fig:local_coordination_fragmentation}
\end{figure}

Figure~\ref{fig:local_coordination_fragmentation} shows that both network groups move toward higher dyadic success over time, but increasing local success does not automatically imply convergence to a single population-wide convention. The trajectories therefore motivate two additional run-level outcomes: \emph{steady-state fragmentation}, defined as the average number of unique conventions in the final 50 rounds of each simulation, and \emph{settling time}, defined as the first round after which the trajectory of unique conventions never again rises above the steady-state fragmentation level by more than 0.5 conventions. Settling time as defined here measures speed of \emph{durable} convergence to a run's own steady state, not speed of approach to global consensus. Because centralized runs often plateau at a higher fragmentation level, faster settling does not imply faster convergence to a single convention.

\begin{figure}[htbp]
    \centering

    \begin{subfigure}[b]{0.45\textwidth}
        \centering
        \includegraphics[width=\linewidth]{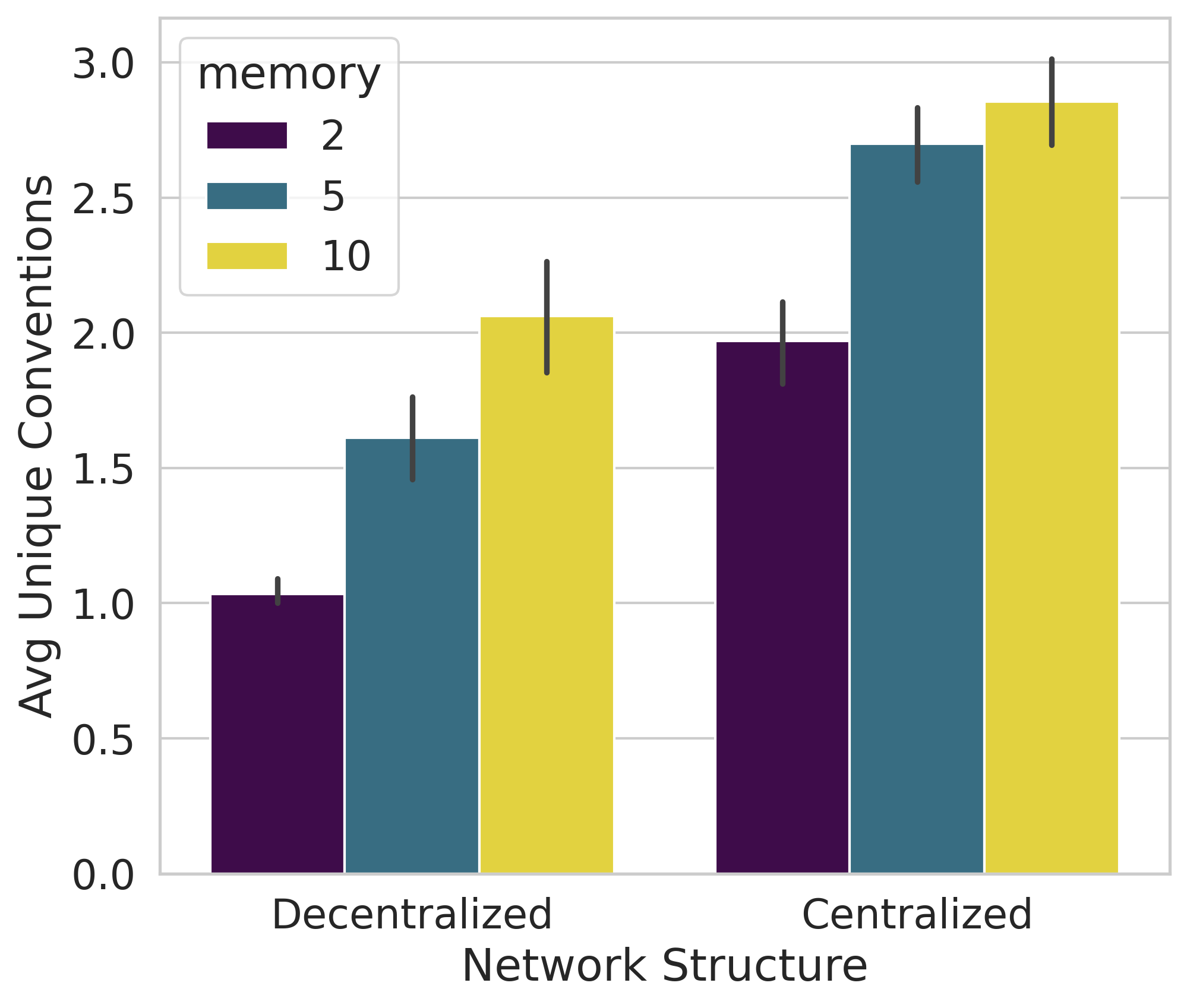}
        \caption{Steady-state fragmentation}
        \label{fig:steady_state}
    \end{subfigure}
    \hfill
    \begin{subfigure}[b]{0.45\textwidth}
        \centering
        \includegraphics[width=\linewidth]{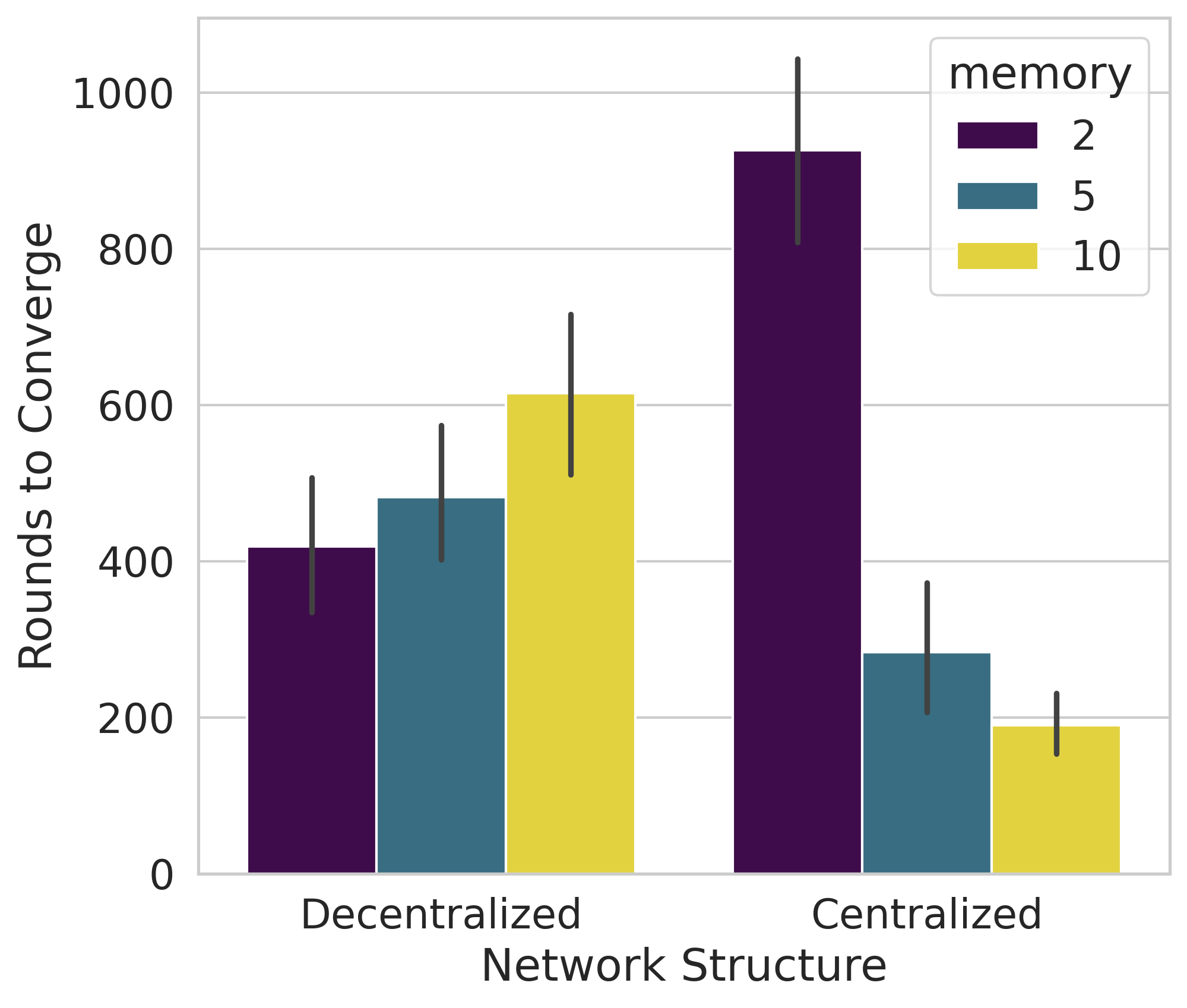}
        \caption{Settling time}
        \label{fig:settling_time}
    \end{subfigure}
    
    \caption{Impact of memory and network structure on convention dynamics. 
    Panel (a) reports the average number of unique conventions in the final $50$ rounds. 
    Panel (b) reports the number of rounds after which the trajectory of unique conventions never again rises above the steady-state fragmentation level by more than 0.5 conventions. 
    Error bars indicate 95\% confidence intervals across independent simulation runs.}
    \label{fig:combined_analysis}
\end{figure}

Figure~\ref{fig:steady_state} shows that longer memory increases steady-state fragmentation in both network types. A two-way ANOVA confirmed significant main effects of memory, $F(2,426)=81.73$, $p<0.001$, and network type, $F(1,426)=225.50$, $p<0.001$, with no significant interaction, $F(2,426)=1.86$, $p=0.156$. Thus, increasing memory consistently increases the number of surviving conventions, while centralized networks preserve more convention diversity than decentralized networks.

Figure~\ref{fig:settling_time} shows a different pattern for settling time (defined above as the round after which the trajectory remains within $0.5$ conventions of its run-specific steady-state). The main effect of memory was significant, $F(2, 426) = 25.64$, $p < 0.001$, $\eta_p^2 = 0.107$, indicating that memory depth affects settling time overall. The main effect of network type was not significant, $F(1, 426) = 1.10$, $p = 0.295$, $\eta_p^2 = 0.003$: averaged across memory levels, centralized and decentralized networks reached their respective steady states in comparable times. Crucially, the interaction between memory and network type was large and highly significant, $F(2, 426) = 57.77$, $p < 0.001$, $\eta_p^2 = 0.213$, indicating that the direction of memory's effect on settling time depends on the network in which agents are embedded.In decentralized networks, settling time \emph{increased} monotonically with memory: longer histories produce longer time to reach steady state. In centralized networks, the relationship reversed: settling time \emph{decreased} sharply with memory. Centralized networks under short memory ($M=2$) settled most slowly of all conditions, plausibly because short memory provides too little partner history to stabilize on a single local convention while clustered neighborhoods supply repeated but conflicting local signals; longer memory in centralized networks accumulates this signal and accelerates settling on locally-reinforced conventions.

Together, Figures~\ref{fig:steady_state} and~\ref{fig:settling_time} reveal a settling-vs-unity trade-off whose direction depends on memory depth. Here 'settling' denotes how quickly a run reaches its own steady state (which may include multiple surviving conventions), while 'unity' denotes whether that steady state is a single convention. Centralized networks consistently preserve more convention diversity than decentralized networks (Figure~\ref{fig:steady_state}), but their settling speed depends sharply on how much agents remember (Figure~\ref{fig:settling_time}). At long memory ($M=10$), centralized networks settle roughly three times faster than decentralized networks (190 vs.\ 615 rounds) but plateau at higher steady-state fragmentation; clustered neighborhoods rapidly reinforce locally agreed conventions while different clusters stabilize on different ones. At short memory ($M=2$), the pattern reverses: decentralized networks settle more than twice as fast as centralized ones (419 vs.\ 926 rounds) and reach cleaner global consensus, because short-memory agents in clustered topologies cannot accumulate enough partner history to stabilize a single neighborhood convention. We verify in Appendix~\ref{app:settling_robustness} that this sign-reversal is robust to the convergence tolerance: the direction of memory's effect within each group and the $M=10$ crossover are preserved across tolerances spanning an order of magnitude around the reported value, and under a plateau-normalized band.

The settling-vs-unity trade-off is therefore not a fixed property of network topology; it is the joint product of topology and memory depth. The mechanism resembles a phenomenon documented in different domains of social simulation: when agents weight prior outcomes heavily, local success can reinforce within-cluster homogeneity at the cost of population-level diversity, contributing to polarization or fragmentation~\cite{horn2026success}.

\subsection{Structural Position and Coordination Resilience}
\label{sec:structural_resilience}

We next examine whether an agent's position \emph{within its own network} affects its individual coordination success. Because betweenness and clustering vary both within and across our eight topologies, a pooled scatter of agent score against either metric would conflate within-network agent-level effects with between-network topology-level differences. We therefore center each metric within network and fit linear mixed-effects models with random intercepts for network identity~\cite{bates2015fitting}. The within-network coefficients describe how the position of an agent \emph{relative to its peers in the same topology} predicts coordination success, holding network structure and run constant. Two local features are especially relevant: betweenness centrality, which captures brokerage across otherwise separated parts of the network, and local clustering, which captures how densely connected an agent's neighbors are. Together, these measures distinguish agents exposed to conflicting cross-cluster signals from agents embedded in locally reinforcing neighborhoods.

\begin{figure}[htbp]
    \centering
    \begin{subfigure}[t]{0.49\textwidth}
        \centering
        \includegraphics[width=\textwidth]{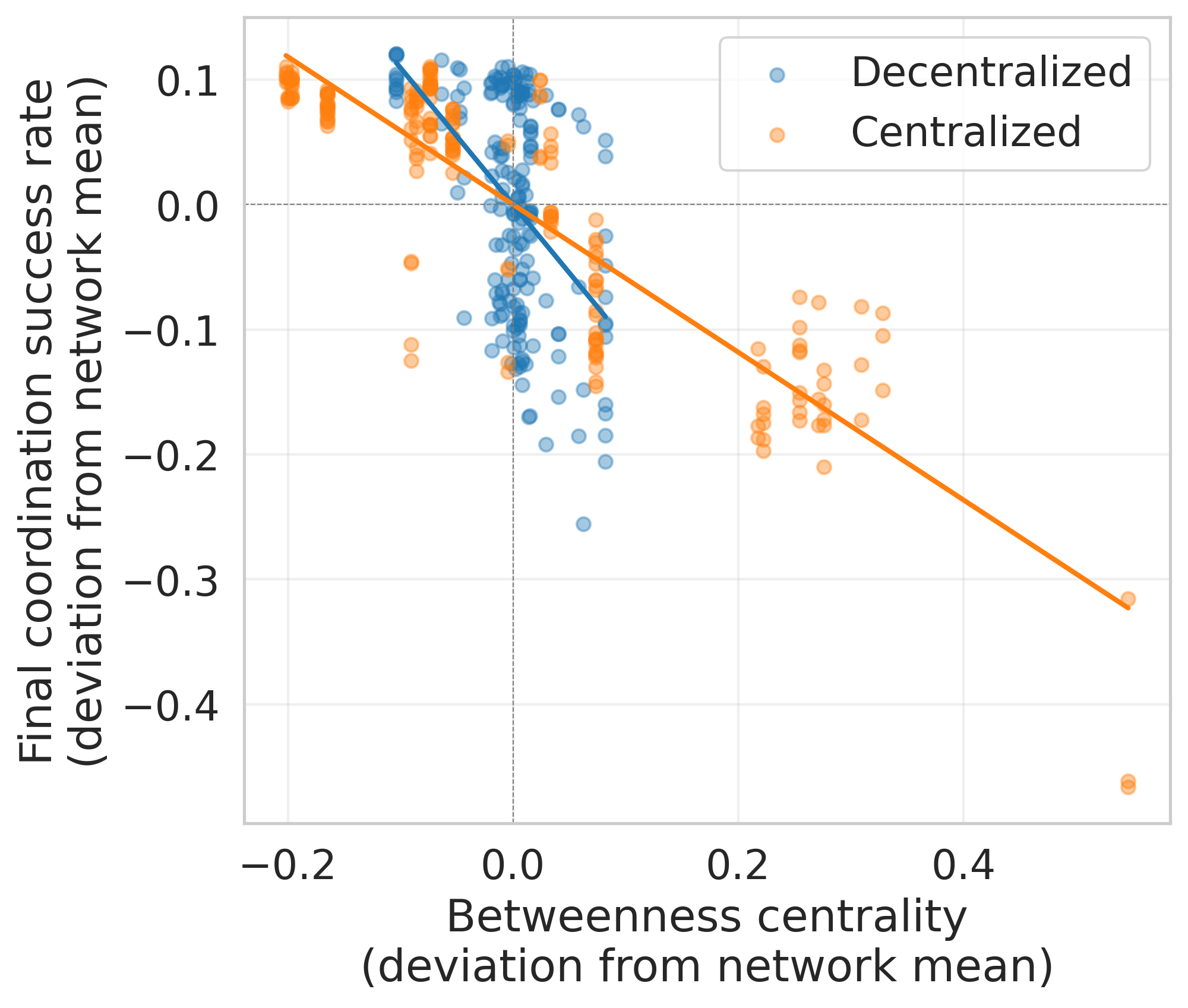}
        \caption{Brokerage effect, within network}
        \label{fig:betweenness_performance}
    \end{subfigure}
    \hfill
    \begin{subfigure}[t]{0.49\textwidth}
        \centering
        \includegraphics[width=\textwidth]{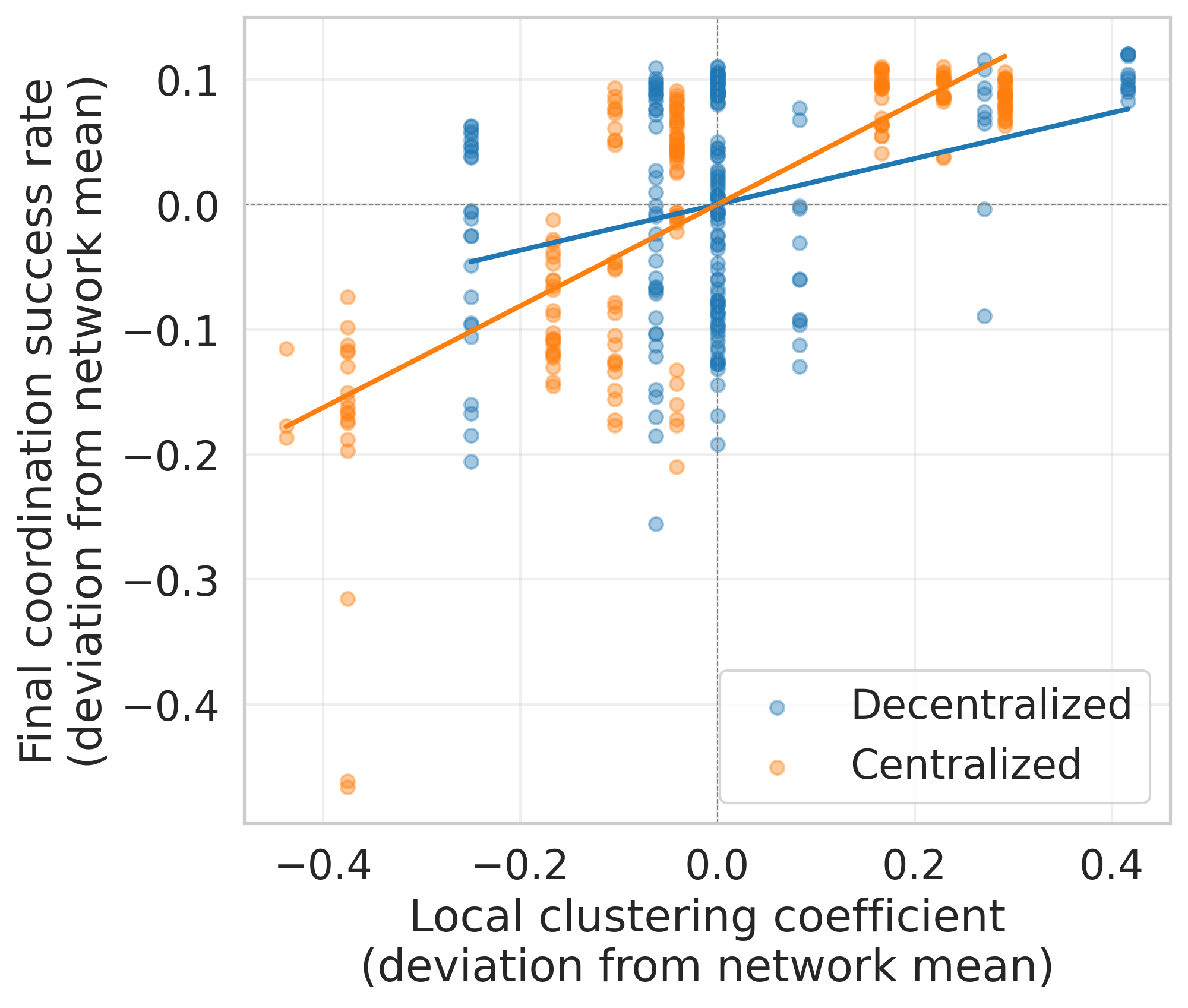}
        \caption{Clustering effect, within network}
        \label{fig:clustering_performance}
    \end{subfigure}
    \caption{Agent coordination success by structural position, within network. 
    Both axes in each panel report deviations from each agent's network mean, isolating within-topology variation in agent position. 
    Panel (a) plots within-network deviation in coordination success rate against within-network deviation in betweenness centrality; high-betweenness agents (right of zero) occupy brokerage positions between subgroups of their topology. 
    Panel (b) plots the same outcome against within-network deviation in local clustering coefficient. 
    The decentralized slope in panel (b) is shallower because two of the four decentralized topologies (A and B) have zero clustering for every node and thus contribute no within-network clustering variation. 
    Solid lines show within-group linear fits.}
    \label{fig:structural_resilience}
\end{figure}

Figure~\ref{fig:structural_resilience}a shows a brokerage penalty: agents with higher betweenness centrality \emph{relative to their peers in the same network} achieve lower final coordination success rates. The within-network mixed-effects regression confirms this across all memory conditions, with the strongest penalty under long memory and a significantly weaker penalty under short memory. The implied within-network slopes are $\beta = -0.49$ at $M = 2$, $\beta = -0.69$ at $M = 5$, and $\beta = -0.67$ at $M = 10$. Formally, the M=10 reference slope is significant ($z = -32.2$, $p < 0.001$); the M=2 slope differs significantly from M=10 ($\beta_{\text{interaction}} = +0.177$, $z = 6.0$, $p < 0.001$), while the M=5 slope does not ($\beta_{\text{interaction}} = -0.018$, $z = -0.6$, $p = 0.55$). The negative slope appears in every one of the eight topologies separately, so the penalty is a property of agent position within a topology, not an artifact of which topology one is comparing across. The mechanism is consistent across memory levels: bridge agents are more likely to receive conflicting signals from different neighborhoods, and longer memory preserves more of those incompatible past interactions, making it harder to stabilize on a single convention.

Figure~\ref{fig:structural_resilience}b shows the complementary role of local clustering. Within each network, agents embedded in more densely clustered neighborhoods achieve higher coordination success across all memory conditions, with the benefit growing monotonically in memory depth. The implied within-network slopes are $\beta = +0.24$ at $M = 2$, $\beta = +0.35$ at $M = 5$, and $\beta = +0.39$ at $M = 10$. Both interaction terms differ significantly from the M=10 reference (M=2: $\beta_{\text{interaction}} = -0.155$, $p < 0.001$; M=5: $\beta_{\text{interaction}} = -0.042$, $p = 0.04$). Each of the six networks containing within-topology clustering variation shows a positive slope. This is consistent with the idea that dense neighborhoods act as convention incubators, where repeated exposure to the same local partners reinforces stable conventions and buffers agents against competing global signals. Longer memory amplifies the protective value of clustering, because clustered neighborhoods supply the same conventions repeatedly and longer memory accumulates that consistent signal. Short memory leaves clustering's protection partly unrealized.

\subsection{Cross-Model Agreement Check}
\label{sec:crossmodel}

To ensure that the observations from our primary simulations are not artifacts of the specific LLM employed (\textit{Gemini 2.0 Flash}), we conducted a cross-validation experiment using two alternative model families: \textit{OpenAI GPT-4o-mini} and \textit{Anthropic Claude Haiku 4.5}. Rather than re-running the full simulations, we adopted a cross-examination approach. We extracted interactions from Rounds 100--200, a high-entropy interval of relatively low consensus, and applied stratified random sampling to select $n=10$ histories per combination of network topology and memory size. This produced a test set of $240$ decision points. Each sampled history, containing the agent's memory of past choices and outcomes, was presented to the alternative models using the same system prompt as in the main simulation. We then compared each model's selected convention to the original \textit{Gemini Flash 2.0} choice for the same history. This analysis measures cross-model agreement rather than accuracy, since the original \textit{Gemini Flash 2.0} choice is a reference decision rather than ground truth.

Agreement rates were uniformly high and far above the 10\% chance baseline: \textit{Claude Haiku 4.5} matched the original \textit{Gemini} choice in 95.4\% of the $240$ sampled decisions (95\% CI roughly $\pm 2.6\%$), and \textit{GPT-4o-mini} matched in 94.2\% (95\% CI roughly $\pm 3.0\%$); see Appendix~\ref{app:cross_model_agreement} for the full breakdown. This suggests that the observed decision patterns are not purely idiosyncratic to the primary model, although a full robustness test would require rerunning the complete simulations with each model.

\section{Conclusions}
\label{sec:conclusion}

The most striking finding is that memory's effect on the time to reach steady state is not just modulated by network structure; it is reversed by it. Longer memory increases settling time in decentralized topologies and decreases it in centralized ones. This is not a quantitative moderation but a directional reversal in how quickly populations lock in to their long-run state, and it has direct implications for how multi-agent LLM systems should be designed. We emphasize that 'settling faster' in centralized topologies means locking in to a more fragmented plateau, not converging to a single shared convention.

The results identify a memory-mediated trade-off between local efficiency and global unity. Centralized topologies act as convention incubators \emph{when memory is long enough}: agents with $M=5$ or $M=10$ rapidly stabilize on locally reinforced conventions, but this speed preserves global fragmentation when different clusters stabilize around different conventions. With short memory, the same centralized structures settle slowest of any condition, because agents lack the partner history needed to commit to a single local convention. Decentralized networks instead expose agents to a more diverse interaction environment; this slows settling at long memory but accelerates it at short memory, and in either case makes local conventions less insulated, increasing the likelihood of system-wide agreement.

The broader lesson that follows is that local coordination is not a proxy for global consensus. High dyadic success can coexist with persistent fragmentation, which means evaluation metrics focused on pairwise interaction quality can mask systemic disagreement at the population level.

Memory's role is therefore not the simple 'inertia' story familiar from prior work on cooperation. Longer memory can preserve obsolete or locally conflicting information, hindering adaptation in decentralized topologies; but in clustered, centralized topologies, longer memory accumulates the consistent local signal that short memory cannot, accelerating settling. Short memory is not uniformly an asset and long memory is not uniformly a liability: each carries opposite consequences in the two structural regimes.

Unlike prior findings in human cooperation experiments, where intermediate memory windows often optimize coordination, the LLM-agent dynamics observed here are closer to monotonic: Within the memory range we tested, longer memory consistently increases surviving conventions. This contrast supports treating LLM-agent populations as a system of interest in their own right rather than as proxies for human behavior. Notably, the effect of memory on settling time reverses sign across topologies: longer memory slows settling in decentralized networks (419 to 615 rounds from $M=2$ to $M=10$) but sharply accelerates it in centralized ones (926 to 190 rounds across the same range). The same parameter change therefore pushes systems in opposite directions depending on the structure in which agents are embedded, and the magnitude of this reversal is large: at $M=10$, centralized networks settle roughly three times faster than decentralized ones, while at $M=2$ the ordering flips. 

An agent's specific location within its topology also matters, and the effect is robust within every network we tested. High-betweenness bridge agents experience a brokerage penalty across all memory conditions ($\beta = -0.49$ at $M=2$, $\beta = -0.69$ at $M=5$, $\beta = -0.67$ at $M=10$), and the negative slope holds within all eight topologies separately. Agents embedded in more densely clustered neighborhoods achieve more stable coordination across memory conditions (within-network slope ranging from $\beta = +0.24$ at $M=2$ to $\beta = +0.39$ at $M=10$, holding within all six networks containing within-topology clustering variation), consistent with the role of dense local neighborhoods as convention incubators. Memory modulates both effects: longer memory deepens the brokerage penalty for bridge agents, and longer memory amplifies the protective benefit of clustering for embedded agents. The same memory horizon that hurts an agent at a structural bridge helps an agent in a dense pocket of the same network.

Agents' choices are better described by Fictitious Play (belief about partners) than by Reinforcement Learning (own past payoffs). The advantage is largest early, when histories are still informative; once local coordination stabilizes, the two rules predict nearly identical choices. This points to a practical implication: in coordination settings, shaping what agents know about each other matters more than shaping their rewards.

These findings have direct implications for practical applied multi-agent AI design. Memory depth and communication topology should be treated as engineering parameters rather than background implementation details, and they should be treated jointly, as complementary parameters, not as separate ones. Systems designed for rapid local task execution may benefit from centralized communication \emph{combined with longer memory horizons}, which together accelerate neighborhood-level lock-in but risk persistent fragmentation across clusters. Systems designed for shared standards, common protocols, or collective agreement may instead require more decentralized mixing and shorter memory horizons, which reduce local lock-in and make system-wide consensus more likely. In practical terms, designers of LLM-agent systems should not optimize memory length or communication architecture independently: the same memory capacity can stabilize coordination in one topology while producing inertia and fragmentation in another.

\section*{Limitations and Outlooks}

This study focused on homogeneous populations, whereas real-world systems often involve agents with diverse cognitive capacities. Future research should investigate heterogeneous populations, where agile agents may act as catalysts for convergence and slow-adapting agents serve as anchors of stability, as suggested in prior studies of heterogeneous multi-agent coordination~\cite{centola2007complex, kearns2006experimental}. Furthermore, our agents operated under zero-shot constraints without Chain-of-Thought (CoT) reasoning. While this isolated the effects of memory depth, it excludes the potential for reasoning-enhanced agents to overcome cognitive inertia by rationalizing partner behavior. Such investigations could leverage emerging AI social platforms, like \textit{Moltbook} or \textit{Chirper}, or \textit{Grokipedia}~\cite{yee2026molt,fadaei2026gender,mehdizadeh2025epistemic}, which simulate complex interaction networks among LLM-based agents and allow researchers to study convergence, influence, and information diffusion in more realistic, large-scale environments~\cite{lin2026exploring,luo2023analyzing}.

A natural next step is to study coordination on \emph{co-evolving} networks, where the topology itself is generated and revised by LLM agents as they interact. Recent work shows that LLM-driven agents grow networks through preferential attachment and develop strong homophilic clustering when social attributes are introduced~\cite{mehdizadeh2025homophily}. Combining that endogenous-topology setting with the convention-formation task studied here would let researchers ask whether the memory--topology interaction we document is preserved, amplified, or dissolved when network structure is itself a product of agent behavior.

We restricted our simulations to small, fixed-degree networks ($N=16, k=3$). We acknowledge that scaling properties may alter these dynamics; future work must investigate if fragmentation persists in large-scale, heterogeneous topologies (e.g., scale-free or small-world networks with high mean degree) or if phase transitions emerge as system size increases~\cite{kim2025towards}.

Our analysis centered on the Naming-Game, a pure coordination task with fully aligned incentives. Extending this framework to social dilemmas, such as the Iterated Prisoner’s Dilemma, would allow examination of scenarios where individual self-interest conflicts with collective outcomes~\cite{nowak1992evolutionary}. Similarly, moving beyond static networks to dynamic, rewiring topologies would enable study of the co-evolution of norms and social structure, as observed in both human and artificial social networks~\cite{gross2008adaptive}.

Our agents queried the model with default decoding parameters, which include non-zero sampling temperature. This is consistent with how LLM agents are typically deployed and gives the population a baseline level of stochastic exploration. It also means that observed convergence is not driven solely by deterministic best-response: the same prompt can yield different choices on different calls. A controlled study at temperature 0 (and at several intermediate temperatures) would help disentangle the effect of decoding stochasticity from the effects of memory and topology, and is a natural follow-up.

Future work should also explore complex epistemic tasks, in which LLM agents validate and propagate factual information rather than arbitrary labels. The convention-formation problem we study has a benign property: any of the ten letters is an acceptable equilibrium. In epistemic settings, that symmetry breaks, because some claims are better supported than others. As LLMs increasingly serve as substitutes for traditional knowledge bases, AI-curated encyclopedias such as Grokipedia have already been shown to systematically restructure citation authority, downweighting peer-reviewed academic sources in favor of user-generated and corporate ones~\cite{mehdizadeh2025epistemic}. Whether the speed--unity trade-off documented here generalizes to settings where conventions carry truth-value is a critical open question for the design of robust multi-agent systems~\cite{shin2025automating}.

\section*{Statements and Declarations}

\textbf{Supplementary Information:} Supplementary material relevant to this study is provided in the appendix.

\textbf{Conflict of Interest:} The authors declare no known competing financial interests or personal relationships that could have influenced the work reported in this paper.

\textbf{Data Availability:} The full set of simulation outputs will be deposited on Zenodo upon publication. During peer review, the dataset is available from the corresponding author upon request.

\textbf{Code Availability:} The simulation source code will be posted at \url{https://github.com/AliakbarMehdizadeh/llm-naming-game}. During peer review, the code will be available from the corresponding author upon request.

\textbf{Author Contributions:} AM \& MH contributed equally to the conceptualization and methodology. AM collected the data, carried out the formal analysis, and wrote the article. MH advised, reviewed, and edited the manuscript.

\bibliographystyle{ieeetr}  
\bibliography{references}  

\begin{appendices}
    
\counterwithin{table}{section} 
\counterwithin{figure}{section}
\renewcommand\thesection{S.I.\arabic{section}}

\appendix

\section{Intrinsic Letter Bias in the LLM: First-Round Choices}
\label{sec:first_round_bias}

\begin{figure}[htbp]
    \centering
    \includegraphics[width=0.7\textwidth]{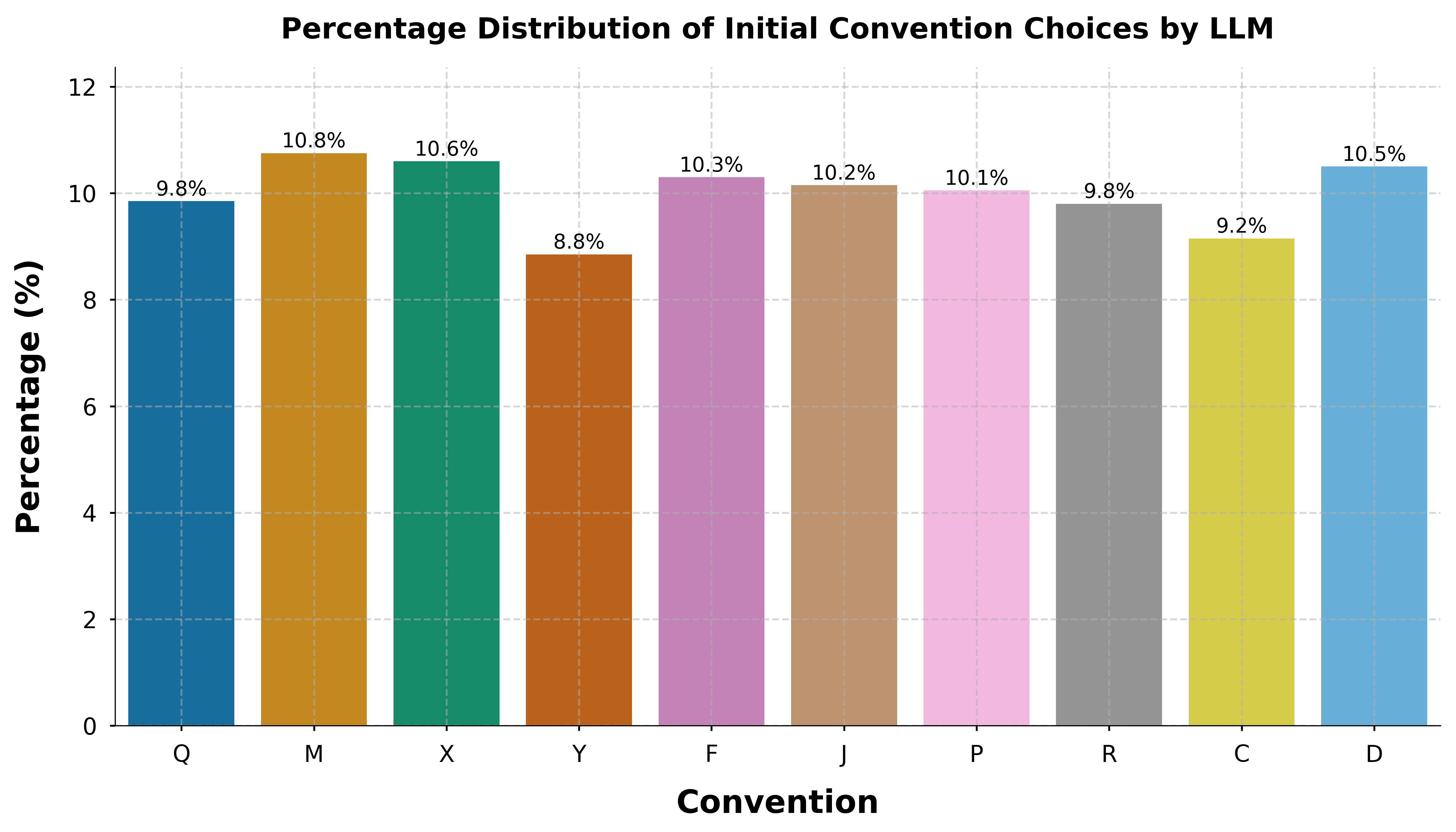}
    \caption{Distribution of initial convention choices made by the \texttt{Gemini 2.0 Flash} model across 2,000 independent runs.}
    \label{fig:gemini_2_0_bias}
\end{figure}

A challenge in studying convention formation is controlling for biases that might favor one arbitrary norm. We selected a 10-letter set \{'C', 'D', 'F', 'J', 'M', 'P', 'Q', 'R', 'X', 'Y'\} shown to be statistically neutral for certain LLM agents. This choice allows us to isolate the effects of \textit{network topology} and \textit{agent memory}, the primary focus of this study. To verify neutrality for our model, we analyzed the first-round choices of \texttt{Gemini 2.0 Flash} across $2,000$ independent simulations with no memory, presenting the conventions in random order. The resulting distribution (Figure~\ref{fig:gemini_2_0_bias}) was nearly uniform. A \texttt{Chi-Squared Goodness-of-Fit} test confirmed no significant intrinsic bias $(\chi^2(9, N=2000) = 4.35, p \approx 0.83)$, ensuring that subsequent dynamics reflect social and network effects rather than pre-existing model preferences.

\section{Supplementary Material: Networks Structural Metrics}
\label{app:network_metrics}

\begin{table}[htbp]
\centering
\caption{Structural properties of the eight network topologies.}
\label{tab:networks_labeled}
\begin{tabular}{clcccccc}
\hline
\textbf{Network Label}& \textbf{Topology} & \textbf{Radius} & \textbf{Diameter} & \textbf{Closeness} & \textbf{Betweenness} & \textbf{Clustering} & \textbf{Constraint} \\
\hline
A & Min avg betweenness & 3 & 3 & 0.45 & 0.09 & 0.00 & 0.33 \\
B & Min avg clustering & 3 & 4 & 0.44 & 0.09 & 0.00 & 0.33 \\
C & Max max closeness & 3 & 5 & 0.41 & 0.10 & 0.06 & 0.36 \\
D & Max var constraint & 3 & 6 & 0.39 & 0.12 & 0.25 & 0.47 \\
E & Max avg clustering & 6 & 6 & 0.31 & 0.16 & 0.50 & 0.60 \\
F & Max max betweenness & 3 & 6 & 0.31 & 0.17 & 0.37 & 0.54 \\
G & Min max closeness & 5 & 9 & 0.27 & 0.20 & 0.37 & 0.53 \\
H & Max avg betweenness & 5 & 9 & 0.27 & 0.20 & 0.44 & 0.57 \\
\hline
\end{tabular}
\end{table}

\section{Extended Behavioral Analysis}
\label{appendix:detailed_analysis}

\begin{figure}[htbp]
    \centering
    \includegraphics[width=\textwidth]{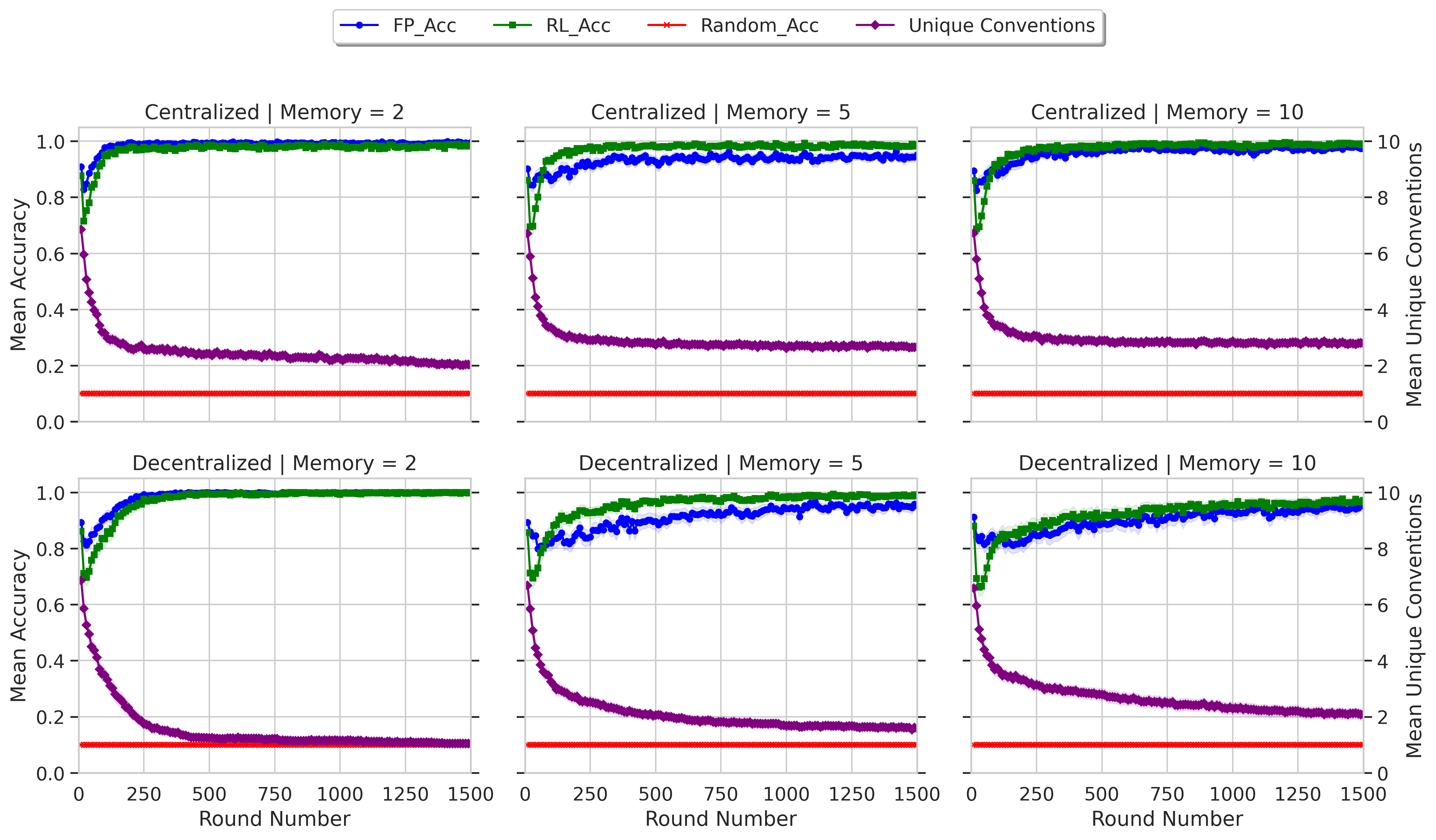} 
    \caption{Faceted Analysis of Strategic Convergence. The grid displays the interaction between network groups (rows) and memory sizes (columns). Each subplot tracks the mean accuracy of the behavioral models (left y-axis) and the mean unique conventions (right y-axis) over 1,600 rounds. Shaded areas represent the 95\% confidence interval.}
    \label{fig:faceted_convergence_details}
\end{figure}

While the main text provides an aggregated view of the simulation results, a more granular examination of the data reveals how specific structural and cognitive constraints influence coordination. Figure~\ref{fig:faceted_convergence_details} presents the evolution of model accuracy and convention convergence, faceted by network topology (Centralized vs. Decentralized) and memory capacity ($M \in \{2, 5, 10\}$).

\section{Cross-Model Agreement: Detailed Results}
\label{app:cross_model_agreement}

This appendix provides the full results of the cross-model agreement check summarized in the main text. The purpose of the analysis is to assess whether the decision patterns observed under the primary model (\textit{Gemini 2.0 Flash}) generalize across LLM families, or whether they reflect idiosyncratic behavior of a single model.

\subsection*{Procedure}

From the full set of 432 primary simulations, we extracted all dyadic interactions occurring in Rounds 100--200. This window was chosen deliberately: it falls after the initial random-exploration phase but before most runs have reached full consensus, so it concentrates decisions on \emph{informative} histories where the agent must actively weigh competing conventions rather than simply repeat an established equilibrium choice.

Within this window, we applied stratified random sampling to select $n=10$ interaction histories per (network, memory) cell, yielding a balanced test set of $8 \times 3 \times 10 = 240$ decision points.

Each sampled history was replayed to two alternative LLMs --- \textit{OpenAI GPT-4o-mini} and \textit{Anthropic Claude Haiku 4.5} --- using the same system prompt, user prompt template, and shuffled conventions list as in the main simulation. Both alternative models were queried with default decoding parameters matching those of the primary model. We then compared each alternative model's selected convention to the original \textit{Gemini} choice for the identical history. Agreement is defined as exact match of the selected convention letter. Under uniform random choice over $K=10$ available conventions, the chance-level agreement rate is $10\%$.

\subsection*{Aggregate agreement}

Both alternative models agreed with the primary model on the vast majority of decisions, well above chance:

\begin{itemize}
    \item \textbf{Claude Haiku 4.5}: 229/240 = 95.4\% agreement (95\% CI: [92.8\%, 98.0\%]).
    \item \textbf{GPT-4o-mini}: 226/240 = 94.2\% agreement (95\% CI: [91.2\%, 97.2\%]).
\end{itemize}

Both rates are roughly $9\times$ the chance baseline, and their confidence intervals overlap substantially, indicating comparable agreement levels across providers.

\begin{figure}[htbp]
    \centering
    \includegraphics[width=0.6\textwidth]{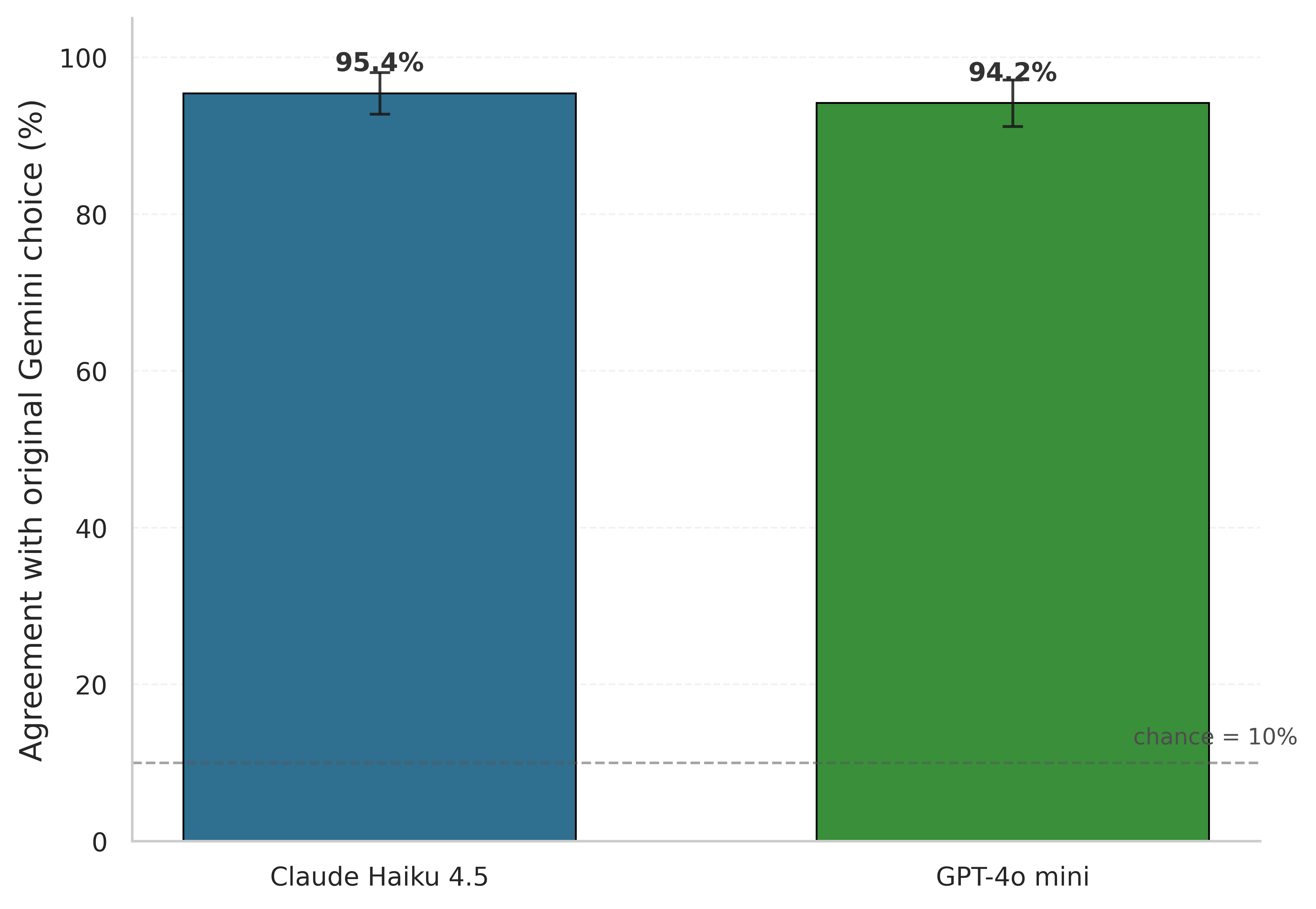}
    \caption{Cross-model decision agreement. Bars show the percentage of instances where each alternative LLM matched the primary model (\textit{Gemini 2.0 Flash}) on the same interaction history ($N=240$ sampled decisions per model). Error bars represent 95\% confidence intervals for a binomial proportion. The dashed line marks chance-level agreement (10\%) given ten available conventions.}
    \label{fig:model_agreement}
\end{figure}

\subsection*{Interpretation}

Agreement above 94\% across two independently developed model families suggests that the convention-choice behavior observed under \textit{Gemini} is not an artifact of a specific provider's training data, alignment procedure, or decoding settings. The approximately 5\% residual disagreement is consistent with two sources: (i) stochasticity in sampling under default decoding parameters; and (ii) genuinely ambiguous histories where multiple conventions have similar support in memory.

We emphasize that this analysis tests \emph{decision-level} agreement, not full behavioral equivalence. A complete robustness check would require rerunning all 432 simulations end-to-end under each alternative model and comparing emergent population-level outcomes. The present analysis provides a computationally efficient robustness check suggesting that the qualitative findings are unlikely to be model-specific. 

\section{Robustness of Settling Time to the Convergence Tolerance}
\label{app:settling_robustness}

Settling time is a derived metric: it depends on a band tolerance $\delta$, the number of conventions by which a run's trajectory may exceed its own steady-state level before we no longer count it as settled (the main text uses $\delta = 0.5$ conventions). Because the headline sign-reversal of memory's effect rests on this metric, we verify that the result is a property of the dynamics rather than of the particular tolerance chosen.

We recomputed settling time for every run across a grid of absolute tolerances, $\delta \in \{0.25, 0.5, 0.75, 1.0, 1.5, 2.0\}$ conventions, holding the steady-state window ($50$ final rounds) and all other settings fixed. Two features of Table~\ref{tab:settling_tol} are mechanical rather than substantive.
First, larger tolerances shorten all settling times, since a wider band is easier to enter. Second, adjacent tolerances sometimes yield identical settling rounds because the convention trajectory is sampled on a $20$-round grid, so the last band crossing moves only when it jumps a grid point.

\begin{figure}[htbp]
    \centering
    \includegraphics[width=\textwidth]{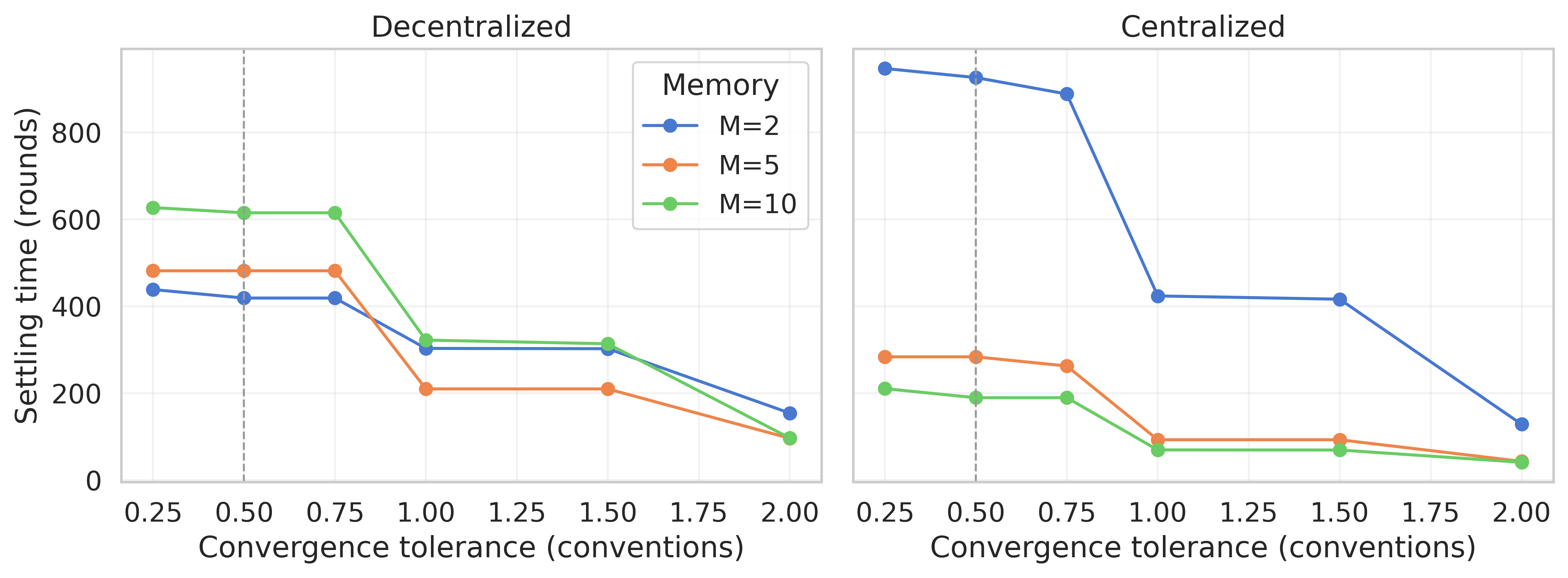}
    \caption{Settling time as a function of the convergence tolerance $\delta$, by memory depth, for decentralized (left) and centralized (right) networks. The dashed line marks the $\delta = 0.5$ value used in the main text. For $\delta \le 1.5$ the memory orderings within each panel do not cross: longer memory consistently slows settling in decentralized networks and speeds it in centralized networks. The curves compress toward small values only at the widest band, $\delta = 2.0$, where the metric degenerates.}
    \label{fig:settling_tol}
\end{figure}

The substantive claims are directional and survive across the reasonable range $\delta \in [0.25, 1.5]$ (Table~\ref{tab:settling_tol} and Figure~\ref{fig:settling_tol}): in decentralized networks settling time \emph{increases} with memory, in centralized networks it \emph{decreases} with
memory, and at long memory ($M=10$) centralized networks settle faster than decentralized networks. At the extreme tolerance $\delta = 2.0$ the metric degenerates: a two-convention band sits roughly twice as high as the decentralized steady-state level (which is close to a single convention), so almost every window already lies inside the band and the settling estimate loses its meaning. Within the non-degenerate range, the reversal is not an artifact of
the $\delta = 0.5$ choice.

\begin{table}[htbp]
\centering
\begin{tabular}{llcccccc}
\toprule
\textbf{Group} & \textbf{Memory} & $\delta{=}0.25$ & $\delta{=}0.5$ & $\delta{=}0.75$ & $\delta{=}1.0$ & $\delta{=}1.5$ & $\delta{=}2.0$ \\
\midrule
\multirow{3}{*}{Decentralized} & $M{=}2$  & 438 & 419 & 419 & 303 & 302 & 154 \\
                               & $M{=}5$  & 482 & 482 & 482 & 210 & 210 &  96 \\
                               & $M{=}10$ & 627 & 615 & 615 & 322 & 314 &  97 \\
\midrule
\multirow{3}{*}{Centralized}   & $M{=}2$  & 947 & 926 & 888 & 424 & 416 & 129 \\
                               & $M{=}5$  & 284 & 284 & 263 &  93 &  93 &  43 \\
                               & $M{=}10$ & 211 & 190 & 190 &  70 &  70 &  41 \\
\bottomrule
\end{tabular}
\caption{Mean settling time (rounds) by network group and memory across convergence tolerances $\delta$. The $\delta=0.5$ column reproduces the values reported in the main text. For $\delta \le 1.5$ the direction of memory's effect (increasing for decentralized, decreasing for centralized) and the $M=10$ ordering (centralized $<$ decentralized) are preserved.}
\label{tab:settling_tol}
\end{table}

\end{appendices}

\end{document}